\documentclass[%
 reprint,
 noeprint,  %
superscriptaddress,
 amsmath,amssymb,
 aps,
]{revtex4-2}

\usepackage{graphicx}%
\usepackage{dcolumn}%
\usepackage{bm}%

\usepackage[caption=false]{subfig}

\usepackage{siunitx}
\DeclareSIUnit\ions{ions}
\DeclareSIUnit\ipsn{\ions \per \nm\squared}
\usepackage[version=4]{mhchem}
\usepackage{booktabs}

\usepackage[disable]{todonotes}

\usepackage{xcolor}
\definecolor{lightblue}{cmyk}{0.230667, 0.076, 0., 0.0706667}    %
\definecolor{darkred}{rgb}{0.717647,   0.086275,   0.086275}    %
\definecolor{darkred}{rgb}{0.85882   0.10196   0.10196}    %

\usepackage{hyperref}
\hypersetup{
    unicode=false,
    pdftoolbar=true,
    pdfmenubar=true,
    pdffitwindow=false,
    pdfstartview={FitH},
    pdftitle={Current-Crowding-Free Superconducting Nanowire Single-Photon Detectors},
    pdfauthor={Stefan Strohauer},
    pdfcreator={Latex with hyperref},
    pdfproducer={pdflatex},
    pdfkeywords={SSPD} {SNSPD} {current crowding} {He ion irradiation} {superconducting thin film}  {radiation damage} {He ion irradiation},
    pdfnewwindow=true,
    pdfborder={0 0 0.5},
    colorlinks=false,
    linkcolor=red,
    citecolor=green,
    filecolor=magenta,
    urlcolor=cyan,
    allbordercolors={lightblue},
}
\usepackage[capitalize,noabbrev,nameinlink]{cleveref}

\makeatletter
\newcommand{\spx}[1]{%
  \if\relax\detokenize{#1}\relax
    \expandafter\@gobble
  \else
    \expandafter\@firstofone
  \fi
  {^{#1}}%
}
\makeatother

\newcommand{\genericdel}[4]{%
  \ifcase#3\relax
  \ifx#1.\else#1\fi#4\ifx#2.\else#2\fi\or
  \bigl#1#4\bigr#2\or
  \Bigl#1#4\Bigr#2\or
  \biggl#1#4\biggr#2\or
  \Biggl#1#4\Biggr#2\else
  \left#1#4\right#2\fi
}

\makeatletter
\newcommand\thefontsize{The current font size is: \f@size pt}

\usepackage{xspace}    %
\newcommand{\eg}{e.\,g.\xspace}

\newcommand{\sspd}{SNSPD\xspace}
\newcommand{\sspds}{SNSPDs\xspace}

\newcommand{\nbtin}{\ce{NbTiN}\xspace}

\newcommand{\Isw}{I_\mathrm{sw}\xspace}

\newcommand{\Ib}{I_\mathrm{b}\xspace}

\newcommand{\Tc}{T_\mathrm{c}\xspace}

\newcommand{\vs}{versus\xspace}

\begin{document}

\title{Current-Crowding-Free Superconducting Nanowire Single-Photon Detectors via Local Helium Ion Irradiation}
\title{Current-Crowding-Free Superconducting Nanowire Single-Photon Detectors
}

\author{Stefan Strohauer}
\email{stefan.strohauer@tum.de}
\affiliation{Walter Schottky Institute, Technical University of Munich,
85748 Garching, Germany
}
\affiliation{TUM School of Natural Sciences, Technical University of Munich,
85748 Garching, Germany}
\author{Fabian Wietschorke}
\author{Christian Schmid}
\affiliation{Walter Schottky Institute, Technical University of Munich,
85748 Garching, Germany
}
\affiliation{TUM School of Computation, Information and Technology, Technical University of Munich,
80333 Munich, Germany
}
\author{Stefanie Grotowski}
\affiliation{Walter Schottky Institute, Technical University of Munich,
85748 Garching, Germany
}
\affiliation{TUM School of Natural Sciences, Technical University of Munich,
85748 Garching, Germany}
\author{Lucio Zugliani}
\affiliation{Walter Schottky Institute, Technical University of Munich,
85748 Garching, Germany
}
\affiliation{TUM School of Computation, Information and Technology, Technical University of Munich,
80333 Munich, Germany
}
\author{Björn Jonas}
\affiliation{Walter Schottky Institute, Technical University of Munich,
85748 Garching, Germany
}
\affiliation{TUM School of Computation, Information and Technology, Technical University of Munich,
80333 Munich, Germany
}
\author{Kai Müller}
\affiliation{Walter Schottky Institute, Technical University of Munich,
85748 Garching, Germany
}
\affiliation{TUM School of Computation, Information and Technology, Technical University of Munich,
80333 Munich, Germany
}
\affiliation{Munich Center for Quantum Science and Technology (MCQST),
80799 Munich, Germany}
\author{Jonathan J. Finley}
\email{jj.finley@tum.de}
\affiliation{Walter Schottky Institute, Technical University of Munich,
85748 Garching, Germany
}
\affiliation{TUM School of Natural Sciences, Technical University of Munich,
85748 Garching, Germany}
\affiliation{Munich Center for Quantum Science and Technology (MCQST),
80799 Munich, Germany}

\date{July 18, 2024}

\begin{abstract}
Detecting single photons is essential for applications such as dark matter detection, quantum science and technology, and biomedical imaging. Superconducting nanowire single-photon detectors (\sspds) excel in this task due to their near-unity detection efficiency, sub-\unit{Hz} dark count rates, and picosecond timing jitter. However, a local increase of current density (current crowding) in the bends of meander-shaped \sspds limits these performance metrics. By locally irradiating the straight segments of \sspds with helium ions while leaving the bends unirradiated, we realize current-crowding-free \sspds with simultaneously enhanced sensitivity: 
after irradiation with \qty{800}{\ipsn}, locally irradiated \sspds showed a relative saturation plateau width of \qty{37}{\percent} while fully irradiated \sspds reached only \qty{10}{\percent}.
This larger relative plateau width allows operation at lower relative bias currents, thereby reducing the dark count rate while still detecting single photons efficiently. We achieve an internal detection efficiency of \qty{94}{\percent} for a  wavelength of \qty{780}{\nm} with a dark count rate of \qty{7}{\mHz} near the onset of saturating detection efficiency.

\end{abstract}


\maketitle

Superconducting nanowire single-photon detectors (\sspds) \cite{Goltsman2001} are widely used across fields such as quantum science and technology \cite{Takesue2007, Chen2022, Liu2023, Bussieres2014, Takesue2015, Shibata2014, Valivarthi2016, Natarajan2012}, astronomy \cite{Wollman2021}, optical communication \cite{Grein2015, Biswas2017}, biology \cite{Xia2021, Tamimi2023}, and medicine \cite{Ozana2021}.
Their ability to detect single photons with near-unity efficiency \cite{Marsili2012a, Korneev2012}, sub-\unit{\Hz} dark count rate \cite{Shibata2015}, picosecond timing jitter \cite{Korzh2020}, and nanosecond reset time \cite{Cherednichenko2021} makes them ideally suited for demanding applications such as photonic quantum computing \cite{Slussarenko2019, Gyger2021, Ferrari2018, Sprengers2011, Reithmaier2013, Reithmaier2015, Majety2023},
particle and dark matter detection \cite{Polakovic2020, Shigefuji2023, Hochberg2019, Chiles2022},
or infrared fluorescence microscopy for in vivo deep brain imaging \cite{Xia2021,Tamimi2023}.
One strategy to further enhance detection efficiency, dark count rate, and timing jitter of 
\sspds is to reduce the effect of current crowding in their 
bends.
Current crowding locally increases the current density in the bends, limiting the maximum applicable bias current through the detector, and thus limiting also the maximum achievable detection efficiency, dark count rate, and timing jitter \cite{Clem2011, Jonsson2022}.
In literature, methods to reduce the effect of current crowding consist of optimized bend geometries \cite{Clem2011, Akhlaghi2012, Henrich2013, Charaev2017, Jonsson2022}, or increased superconductor thickness in the bends \cite{Baghdadi2021, Xiong2022}.
The use of optimized bend geometries is either limited to relatively low-fill factor \sspds, which have lower absorption and detection efficiency, 
or requires a spiral or special meander design with significant wire length overhead, which is incompatible with dense \sspd arrays and limits the timing properties of the detector.
In contrast, variable thickness \sspds successfully reduce current crowding also for highly efficient and compact large fill factor \sspds.
In this work, we introduce 
a novel method to obtain not only current-crowding-free \sspds of arbitrary geometry and fill factor
but also to simultaneously enhance their sensitivity.

\section{Current crowding at low temperatures}
Current crowding describes a non-uniform distribution of current density in an electrical conductor and has its origin for example in variations in material properties or device geometry.
In the context of \sspds, current crowding leads to a local increase of the supercurrent density in the \qty{180}{\degree} bends and at kinks or discontinuities of the \sspd.
Once this current density exceeds the critical value for the superconductor, the \sspd switches to the normal conducting state. 
This maximum current defines the switching current $\Isw$.
In other words, current crowding limits $\Isw$ of the \sspd.

\begin{figure}
 \centering
 \includegraphics{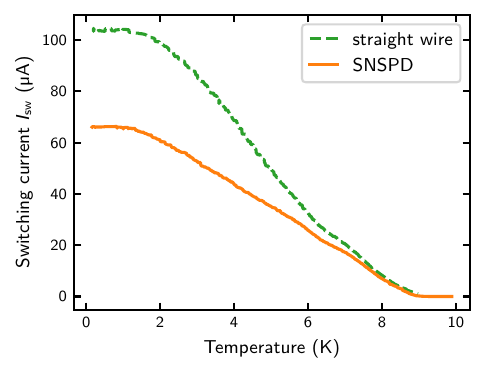}%
     \caption{Switching current \vs temperature for an \sspd and a straight wire. Current crowding causes a reduced switching current for the \sspd over the straight wire, especially towards low temperatures.}
 \label{fig:Isw_vs_T_wire_vs_SSPD_current_crowding}
\end{figure}

It is worthwhile to note that the effect of current crowding is most pronounced at low temperatures ($< 0.7\,\Tc$), while at temperatures close to $\Tc$ the switching current practically coincides for straight and bent wires due to the divergence of the coherence length at temperatures close to the critical temperature \cite{Henrich2012, Hortensius2012}.
To quantify the effect of current crowding for our devices, we measured the temperature dependence of the switching current for \sspds (in meander form with \qty{180}{\degree} bends) and straight wires (without any bend), both consisting of \qty{250}{\nm} wide and \qty{8}{\nm} thick \nbtin wires.
As shown for two representative devices in \cref{fig:Isw_vs_T_wire_vs_SSPD_current_crowding}, the switching current increases and the effect of current crowding becomes more prominent towards low temperatures, reaching a saturating (relative) difference between $\Isw$ of the straight wire and the \sspd at temperatures below \qty{1.6}{\K} (\qty{1}{\K}).
In fact, at temperatures below \qty{1}{\K}, the switching current of \nbtin meander \sspds with a fill factor of \qty{71}{\percent} and \qty{250}{\nm} wide wires is reduced to only \qty{60}{\percent} of $\Isw$ of corresponding straight wires.

At the same time, a high switching current, as close as possible to the limit set by the depairing current density, is desirable because it allows larger plateaus of saturating detection efficiency \cite{Frasca2019,Korzh2020}. 
This makes operation with high detection efficiency at simultaneously low dark count rates possible.
Moreover, high switching currents allow high bias currents through the \sspds that yield high and easily detectable voltage pulses and low timing jitter \cite{Xiong2022}.
Since high switching currents are reached for low temperatures 
where current crowding limits the maximum applicable bias current, a method to avoid current crowding would be highly advantageous.
Moreover, current-crowding-free \sspds are expected to exhibit fewer dark counts since they originate primarily from bends or constrictions with strong current crowding \cite{Akhlaghi2012, Semenov2015, Zhang2014, Baghdadi2021, Zhang2022b}.
Compared to methods presented in literature that solely mitigate current crowding \cite{Clem2011, Akhlaghi2012, Henrich2013, Charaev2017, Jonsson2022, Baghdadi2021, Xiong2022}, the following section introduces a solution to obtain current-crowding-free \sspds of arbitrary geometry that simultaneously exhibit enhanced sensitivity.

\section{Design and concept of current-crowding-free
\sspds with enhanced sensitivity}
\begin{figure*}
 \centering
 \includegraphics{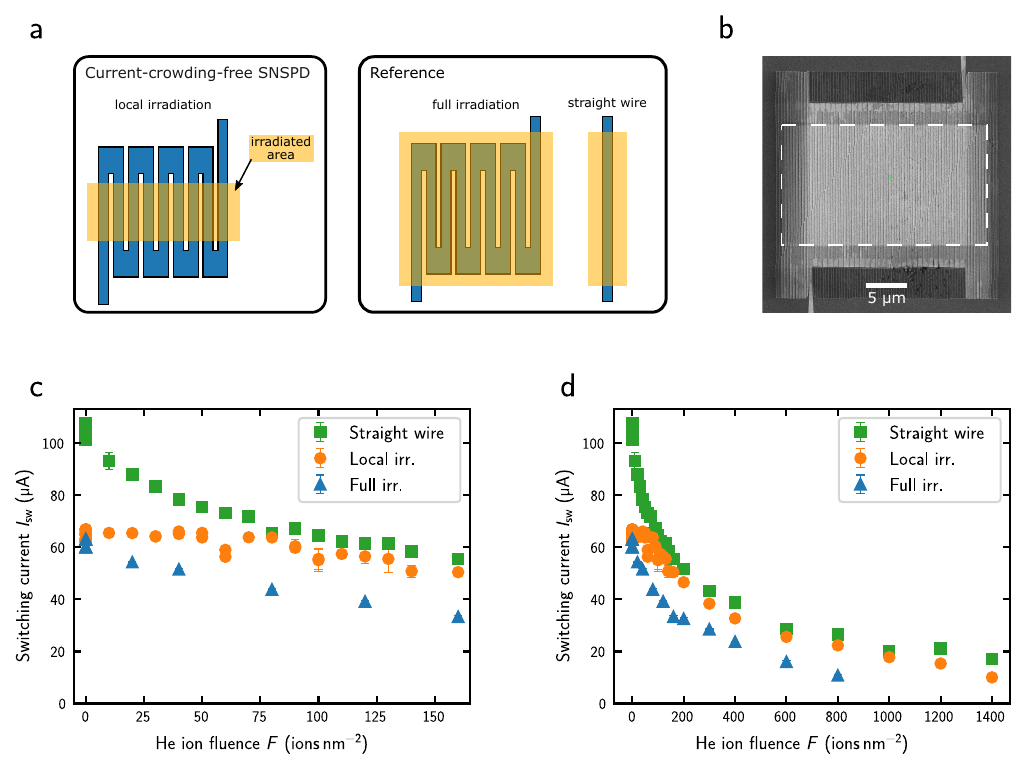}%
     \caption{Current-crowding-free \sspds. 
     \textbf{a}, Local irradiation of the straight detector segments is used to obtain current-crowding-free \sspds. Fully irradiated \sspds and straight wires were fabricated as reference devices.
     \textbf{b}, Scanning electron microscope image of an \sspd after local irradiation. The irradiation area is indicated by the white dashed rectangle. 
     \textbf{c}, \textbf{d}, Switching current \vs He ion fluence for \emph{locally} as well as  \emph{fully} irradiated \sspds and straight wires for the low fluence regime (\textbf{c}) and for the full measured range up a fluence of \qty{1400}{\ipsn} (\textbf{d}). 
}
 \label{fig:Composition_schematic-local-irradiation_SEM_Isw-vs-dose}
\end{figure*}

To demonstrate current-crowding-free \sspds of enhanced sensitivity, we combine the two effects of reduced switching current and enhanced sensitivity of \sspds after helium (He) ion irradiation \cite{Zhang2019,Strohauer2023}.
As discussed in the previous section, the \qty{180}{\degree} bends are the main contribution to current crowding in a meander-type \sspd, thus limiting its switching current.
By locally irradiating only the straight segments of the \sspd while leaving its bends unirradiated, we reduce the critical current density of the straight segments while making them simultaneously more sensitive to single photons.
Once the He ion fluence is large enough, the critical current of the straight segments with homogeneous current density will be lower than the critical current of the bends.
Then, one can expect the overall critical current (or switching current) of the \sspd to be given by the critical current of the straight segments and not being limited by current crowding anymore.
To investigate this hypothesis, three types of devices as shown in \cref{fig:Composition_schematic-local-irradiation_SEM_Isw-vs-dose}a were fabricated on the same chip: 
\sspds that were \emph{locally} (only their straight segments) or \emph{fully} irradiated, as well as \emph{straight wires} without any bend.
They were fabricated from an \qty{8}{\nm} thick \nbtin film on a \ce{Si} substrate with \qty{130}{\nm} thermally grown \ce{SiO2}.
The nominal design of the detectors consists of \qty{250}{\nm} wide nanowires in meander form with \qty{100}{\nm} gaps (fill factor \qty{71}{\percent}) and covers an active area of \qtyproduct[product-units = repeat]{20 x 20}{\um}.
This high fill factor, along with the nominally rectangular design of the bends between neighboring nanowires, was chosen to have a pronounced effect of current crowding.
The reference straight wires were fabricated on the same chip, also with a wire width of \qty{250}{\nm}.
All devices were characterized prior to irradiation by measuring 
their switching current distributions and mean switching currents at a temperature of \qty{1}{\K}.
After characterization of the unirradiated devices, a He ion microscope (Zeiss Orion Nanofab) with an acceleration voltage of \qty{30}{\kV} was used to 
irradiate them with He ions.
An irradiation area of \qtyproduct[product-units = repeat]{25 x 15}{\um} was centered on the \qtyproduct[product-units = repeat]{20 x 20}{\um} detectors to homogeneously expose the straight segments of the nanowires while leaving the bends unirradiated.
In this way, locally irradiated \sspds as shown in \namecrefs{fig:Composition_schematic-local-irradiation_SEM_Isw-vs-dose}~\ref{fig:Composition_schematic-local-irradiation_SEM_Isw-vs-dose}a and \ref{fig:Composition_schematic-local-irradiation_SEM_Isw-vs-dose}b were obtained.
For the reference straight wires, an irradiation area of \qtyproduct[product-units = repeat]{5 x 15}{\um} was chosen.
Moreover, two detectors were fully irradiated (\qtyproduct[product-units = repeat]{25 x 25}{\um} irradiation area).
After irradiation, the devices were characterized again, and for several devices the process of irradiation and subsequent measurement was repeated multiple times.

\section{Reducing current crowding by local He ion irradiation}
As shown in \namecrefs{fig:Composition_schematic-local-irradiation_SEM_Isw-vs-dose}~\ref{fig:Composition_schematic-local-irradiation_SEM_Isw-vs-dose}c and \ref{fig:Composition_schematic-local-irradiation_SEM_Isw-vs-dose}d, the mean switching current of locally irradiated \sspds stays constant for doses up to 
\qty{90+-20}{\ipsn},
while it continuously decreases for fully irradiated \sspds and straight wires.
As anticipated, locally irradiated \sspds exhibit a constant $\Isw$ until the He ion fluence reaches a level that reduces $\Isw$ of the straight wires to comparable values.
Beyond this point, $\Isw$ of locally irradiated \sspds and straight wires follow a similar curve.
The slightly smaller $\Isw$ of locally irradiated \sspds compared to that of the straight wires can be attributed to the higher likelihood of imperfections in one of the many straight segments of an \sspd, unlike in a single straight wire.

\begin{figure}
 \centering
 \includegraphics{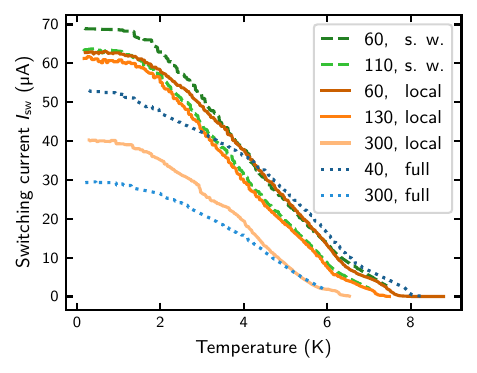}%
     \caption{Switching current $\Isw$ of straight wires (s. w.), locally irradiated \sspds (local), and fully irradiated \sspds (full) \vs temperature.
     The He ion fluence used for irradiation is given by the numbers in front of the device type in units of \unit{\ipsn}.
     }
\label{fig:Isw_vs_T_wire_local_full_and_multiple_doses}
\end{figure}

\Cref{fig:Isw_vs_T_wire_local_full_and_multiple_doses} shows the temperature dependence of the maximum switching current for differently irradiated \sspds and straight wires, determined from the underlying switching current distributions \cite{Brenner2012,Hortensius2012}.
The critical temperature 
decreases continuously with increasing He ion fluence in accordance with literature \cite{Strohauer2023,Zhang2019} and does not depend on device type or irradiation type.
For decreasing temperature, the $\Isw(T)$ curves start to differ and show a smaller slope for fully irradiated \sspds compared to that of locally irradiated \sspds and irradiated straight wires.
Moreover, for similar He ion fluences and low temperatures, $\Isw$ is much smaller for fully irradiated \sspds than for locally irradiated \sspds or straight wires.
For example, at temperatures below \qty{1}{\K} and for a He ion fluence of \qty{300}{\ipsn} it is only \qty{30}{\uA} instead of \qty{40}{\uA}.
While for a  He ion fluence of \qty{60}{\ipsn} the switching current of the straight wire is still higher than that of the locally irradiated \sspd, they almost coincide for the 
two fluences 
\qty{110}{\ipsn} and \qty{130}{\ipsn}.
Those observations 
support
the hypothesis that for small He ion fluences $\Isw$ of the locally irradiated \sspds is not reduced, while it starts to coincide with that of the straight wires for fluences above \qty{90+-20}{\ipsn}.
In this case, the irradiation induced reduction of the switching current in the straight segments is the limiting factor for $\Isw$ instead of current crowding in the bends.
The stronger the effect of current crowding in the unirradiated device, the higher the He ion fluence required for local irradiation such that current crowding no longer plays a role.
Since current crowding limits the switching current of \sspds more as the fill factor increases, this method is particularly useful for high fill factor or micro-scale \sspds, especially if the detector requires He ion irradiation anyway to become single-photon sensitive \cite{Charaev2024,Wang2024}.

\section{Enhanced sensitivity of current-crowding-free \sspds}
\begin{figure*}
 \centering
 \includegraphics{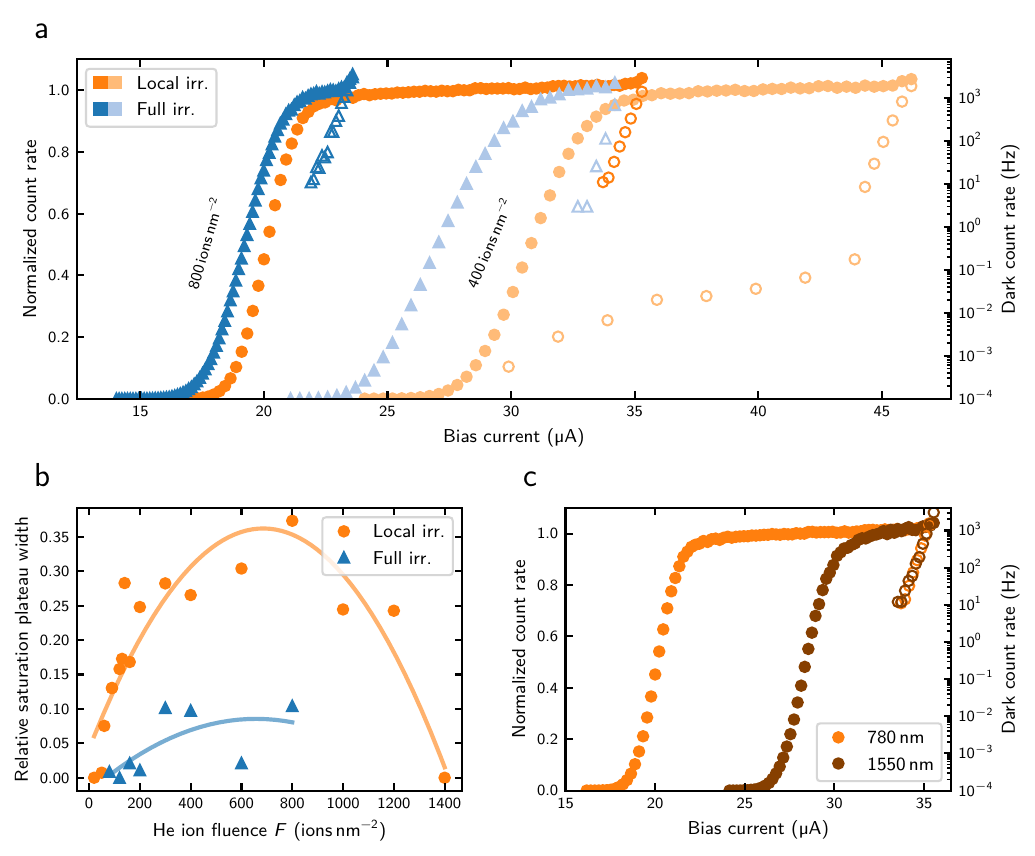}
 \caption{Detection performance of \sspds after local or full irradiation with He ions. 
 \textbf{a}, Normalized count rate (solid symbols) and dark count rate (open symbols) \vs bias current after local (orange) or full (blue) irradiation with He ions. The He ion fluence used for irradiation is shown as annotations at the corresponding count rate curves.
 \textbf{b}, Relative saturation plateau width 
 \vs He ion fluence after local or full irradiation, calculated according to \cref{eq:RSPW}. The solid lines serve as a guide to the eye.
 \textbf{c}, Comparison of the normalized count rate between photons of \qty{780}{\nm} and \qty{1550}{\nm} wavelength for a detector locally irradiated with \qty{800}{\ipsn}.
 All shown dark count rate data was taken such that the integration time per data point was at least 10 times higher than the inverse of the measured dark count rate.}
\label{fig:Composition_CR_RSPW_780vs1550}
\end{figure*}

Helium ion irradiation enhances the sensitivity of \sspds.
However, since we do not irradiate the whole \sspd but intentionally leave the bends unirradiated and, consequently, do not have the irradiation induced decrease of switching current, we expect an enhanced performance due to two effects:
(1) the He ion irradiation induced higher sensitivity, and (2) the higher available bias currents compared to fully irradiated \sspds of the same He ion fluence.
\Cref{fig:Composition_CR_RSPW_780vs1550}a shows a comparison of the normalized count rates between two devices, one irradiated locally and one irradiated fully, both with the same He ion fluences and for photons of \qty{780}{\nm} wavelength.
The locally irradiated \sspd clearly shows higher switching currents and larger saturation plateaus of the count rates, for which the internal detection efficiency is unity.
Moreover, since the dark count rate at a certain bias current $\Ib$ mainly depends exponentially on the ratio $\Ib/\Isw$, 
the dark count rate of locally irradiated \sspds is lower than that of fully irradiated \sspds at similar absolute bias currents or normalized count rates.
Since our \sspds exhibit very low dark count rates, we measured it over an extended period of time for selected devices.
As shown in \cref{fig:Composition_CR_RSPW_780vs1550}a for the device locally irradiated with \qty{400}{\ipsn}, we reach dark count rates of \qty{7}{\mHz} and \qty{0.55}{\mHz} at internal detection efficiencies of \qty{94}{\percent} and \qty{31}{\percent}, respectively.
Also the two regimes of intrinsic dark counts and dark counts originating from black body radiation are clearly visible: the first is given by the strongly current dependent contribution in the high current regime, while the second is the weaker current dependent contribution at lower currents \cite{Kahl2015}.
The black body radiation induced dark count rate could be further suppressed by using cold band-pass filters \cite{Shibata2013}.

In \cref{fig:Composition_CR_RSPW_780vs1550}b we present the relative count rate saturation plateau width, defined as
\begin{equation}
    \sigma_\mathrm{rel} = 
    \frac{I_\mathrm{c} - I_\mathrm{sat}}{I_\mathrm{c}}
    \;,
    \label{eq:RSPW}
\end{equation}
with the critical current $I_\mathrm{c}$, and $I_\mathrm{sat}$ representing the current where the saturation plateau begins (where the normalized count rate reaches 0.95).
The critical current is defined as the smallest bias current where the \sspd shows non-zero resistance.
Since we use a shunt resistor for the CR measurements to prevent latching of the detectors, the \sspds transition to the relaxation oscillation regime at $I_\mathrm{c}$ before switching to the latching state \cite{Kerman2013}.
Most importantly, the relative saturation plateau width is higher for locally irradiated \sspds compared to fully irradiated \sspds of the same He ion fluence.
Furthermore, the relative saturation plateau width increases strongest as a function of He ion fluence for values smaller than \qty{200}{\ipsn} and, as elaborated in the supplementary,  it peaks between \qty{600}{\ipsn} and \qty{1000}{\ipsn} before it decreases again for higher fluences.

To assess the performance of the device locally irradiated with \qty{800}{\ipsn} for optical communication in the C-band, we measured its normalized count rate for photons with a wavelength of \qty{1550}{\nm} and compare it in \cref{fig:Composition_CR_RSPW_780vs1550}c with its performance for \qty{780}{\nm} photons.
For both wavelengths the detector shows saturating internal detection efficiency, even for \qty{1550}{\nm} where the onset of photon detection is shifted to higher bias currents due to the lower sensitivity to longer wavelength photons.
Since it is the same device, the dark count rate is identical for both measurements.

Moreover, we found that the detection pulse height (relevant for assessing the requirements of the readout electronics) of locally irradiated \sspds is higher than that of fully irradiated \sspds of the same fluence, while the pulse decay time (related to the detector's dead time) is similar.
These quantities are further discussed in the supplementary.

\section{Conclusions}
Current crowding strongly limits the maximum bias current, sensitivity, and dark count rate of \sspds at low temperatures.
We demonstrated a method to obtain current-crowding-free \sspds by irradiating the straight segments of \sspds locally with He ions while leaving their bends unirradiated.
Up to a He ion fluence of \qty{90+-20}{\ipsn} the critical current is not reduced, meaning the photon sensitivity is enhanced without sacrificing any of the \sspd's critical current.
Above this threshold, the straight segments of the locally irradiated \sspds show a critical current lower than that of the bends such that current crowding in the bends does not limit the overall critical current anymore.

The boost in performance for locally irradiated \sspds is twofold:
(1) the sensivitiy of the \sspd is enhanced due to He ion irradiation, and (2) the sensitivity is higher than for fully irradiated \sspds due to the higher available bias currents.
Using this approach, we achieved a relative saturation plateau width of \qty{37}{\percent} for a locally irradiated \sspd compared to \qty{10}{\percent} for a fully irradiated \sspd, both irradiated with \qty{800}{\ipsn}.
This larger relative plateau width means that the \sspd can be operated at lower relative bias currents with lower dark count rates and still detect single photons efficiently (\eg, \qty{7}{\mHz} dark count rate at an internal detection efficiency of \qty{94}{\percent}).
Therefore, this method is particularly beneficial for applications that require high detection efficiency combined with low dark count rate, such as quantum key distribution, brain imaging, and dark matter detection.
Furthermore, it is ideally suited for high fill factor or micro-scale \sspds that are prone to current crowding, especially if the detectors require He ion irradiation anyway to become single-photon sensitive \cite{Charaev2024}.
By combining site-selective \cite{Strohauer2023} with local He ion irradiation, one can realize highly and homogeneously performing \sspd arrays.
Moreover, local He ion irradiation can be further used to study the effect of current crowding in different regions of superconducting devices, including \sspds and the influence of current crowding in their bends onto properties such as dark count rate, detection efficiency, and switching current.

\begin{acknowledgments}
We gratefully acknowledge support from the German Federal Ministry of Education and Research (BMBF) via the funding program ``Photonics Research Germany'' (projects MOQUA (13N14846) and MARQUAND (BN106022)) and the funding program ``Quantum technologies -- from basic research to market'' (projects PhotonQ (13N15760), SPINNING (13N16214), QPIS (16K1SQ033), QPIC-1 (13N15855) and SEQT (13N15982)), as well as from the German Research Foundation (DFG) under Germany's Excellence Strategy EXC-2111 (390814868) with the projects PQET (INST 95/1654-1) and MQCL (INST 95/1720-1).
This research is part of the ``Munich Quantum Valley'', which is supported by the Bavarian state government with funds from the ``Hightech Agenda Bayern Plus''.

\end{acknowledgments}

\section*{Methods}
\subsection{Fabrication of \nbtin \sspds}
To fabricate \sspds and straight wires, we deposited an \qty{8}{\nm} thick \nbtin film using DC reactive magnetron sputtering onto a \ce{Si} substrate with a \qty{130}{\nm} thick thermally grown \ce{SiO2} layer.
The superconductor thickness was controlled by measuring the sputtering rate and choosing the sputtering time correspondingly.
Subsequently, we patterned the \sspds and straight wires with electron beam lithography and reactive ion etching, followed by contact pad fabrication using optical lithograpy and gold evaporation \cite{Flaschmann2023}.
The detector design consists of \qty{250}{\nm} wide nanowires in meander form with \qty{100}{\nm} gaps (fill factor \qty{71}{\percent}) and covers an active area of \qtyproduct[product-units = repeat]{20 x 20}{\um}.
To have a pronounced effect of current crowding, a nominally rectangular design of the bends between neighboring nanowires of the \sspds was chosen.
Similar to the detectors, the straight wires were designed to be \qty{250}{\nm} wide and fabricated on the same chip to ensure best comparability.
After characterization of the unirradiated devices, we used a He ion microscope (Zeiss Orion Nanofab) for irradiation with He ions with an acceleration voltage of \qty{30}{\kV}.
For the locally irradiated devices we chose an irradiation area of \qtyproduct[product-units = repeat]{25 x 15}{\um}, centered on the \qtyproduct[product-units = repeat]{20 x 20}{\um} detectors. 
Two detectors were fully irradiated with an irradiation area of \qtyproduct[product-units = repeat]{25 x 25}{\um}.
For the reference straight wires, an irradiation area of \qtyproduct[product-units = repeat]{5 x 15}{\um} was chosen.
After irradiation, we characterized the devices again and repeated the process of irradiation and subsequent measurements for several devices multiple times.

\subsection{Low-temperature measurements}
All detectors were pre-characterized in a cryogenic probe station (Janis) at \qty{4.5}{\K}.
Subsequently, devices exhibiting a similar switching current of \qty{42+-2}{\uA} at \qty{4.5}{\K} were selected for wire bonding and further characterization at temperatures down to \qty{150}{\milli\K} in a closed cycle cryostat with adiabatic demagnetization refrigeration (kiutra GmbH).
Further characterization consisted of  measuring the switching current of each device 10\,--\,250 times to determine a switching current distribution and its mean switching current at a temperature of \qty{1}{\K}.
The switching current measurements were performed by ramping the bias voltage applied at a \qty{50}{\kilo\ohm} resistor in series to the respective device and determining the current at which the total resistance increases by more than \qty{1}{\kilo\ohm}.
Temperature dependent switching current
measurements were performed at temperature ramp rates of \qty{0.075}{\K/\min}, ensuring that during one bias voltage sweep the temperature change was less than \qty{0.1}{\K}.

\bibliography{HIM_paper_II}%

\begin{thebibliography}{53}%
\makeatletter
\providecommand \@ifxundefined [1]{%
 \@ifx{#1\undefined}
}%
\providecommand \@ifnum [1]{%
 \ifnum #1\expandafter \@firstoftwo
 \else \expandafter \@secondoftwo
 \fi
}%
\providecommand \@ifx [1]{%
 \ifx #1\expandafter \@firstoftwo
 \else \expandafter \@secondoftwo
 \fi
}%
\providecommand \natexlab [1]{#1}%
\providecommand \enquote  [1]{``#1''}%
\providecommand \bibnamefont  [1]{#1}%
\providecommand \bibfnamefont [1]{#1}%
\providecommand \citenamefont [1]{#1}%
\providecommand \href@noop [0]{\@secondoftwo}%
\providecommand \href [0]{\begingroup \@sanitize@url \@href}%
\providecommand \@href[1]{\@@startlink{#1}\@@href}%
\providecommand \@@href[1]{\endgroup#1\@@endlink}%
\providecommand \@sanitize@url [0]{\catcode `\\12\catcode `\$12\catcode `\&12\catcode `\#12\catcode `\^12\catcode `\_12\catcode `\%12\relax}%
\providecommand \@@startlink[1]{}%
\providecommand \@@endlink[0]{}%
\providecommand \url  [0]{\begingroup\@sanitize@url \@url }%
\providecommand \@url [1]{\endgroup\@href {#1}{\urlprefix }}%
\providecommand \urlprefix  [0]{URL }%
\providecommand \Eprint [0]{\href }%
\providecommand \doibase [0]{https://doi.org/}%
\providecommand \selectlanguage [0]{\@gobble}%
\providecommand \bibinfo  [0]{\@secondoftwo}%
\providecommand \bibfield  [0]{\@secondoftwo}%
\providecommand \translation [1]{[#1]}%
\providecommand \BibitemOpen [0]{}%
\providecommand \bibitemStop [0]{}%
\providecommand \bibitemNoStop [0]{.\EOS\space}%
\providecommand \EOS [0]{\spacefactor3000\relax}%
\providecommand \BibitemShut  [1]{\csname bibitem#1\endcsname}%
\let\auto@bib@innerbib\@empty
\bibitem [{\citenamefont {Gol'tsman}\ \emph {et~al.}(2001)\citenamefont {Gol'tsman}, \citenamefont {Okunev}, \citenamefont {Chulkova}, \citenamefont {Lipatov}, \citenamefont {Semenov}, \citenamefont {Smirnov}, \citenamefont {Voronov}, \citenamefont {Dzardanov}, \citenamefont {Williams},\ and\ \citenamefont {Sobolewski}}]{Goltsman2001}%
  \BibitemOpen
  \bibfield  {author} {\bibinfo {author} {\bibfnamefont {G.~N.}\ \bibnamefont {Gol'tsman}}, \bibinfo {author} {\bibfnamefont {O.}~\bibnamefont {Okunev}}, \bibinfo {author} {\bibfnamefont {G.}~\bibnamefont {Chulkova}}, \bibinfo {author} {\bibfnamefont {A.}~\bibnamefont {Lipatov}}, \bibinfo {author} {\bibfnamefont {A.}~\bibnamefont {Semenov}}, \bibinfo {author} {\bibfnamefont {K.}~\bibnamefont {Smirnov}}, \bibinfo {author} {\bibfnamefont {B.}~\bibnamefont {Voronov}}, \bibinfo {author} {\bibfnamefont {A.}~\bibnamefont {Dzardanov}}, \bibinfo {author} {\bibfnamefont {C.}~\bibnamefont {Williams}},\ and\ \bibinfo {author} {\bibfnamefont {R.}~\bibnamefont {Sobolewski}},\ }\bibfield  {title} {\bibinfo {title} {Picosecond superconducting single-photon optical detector},\ }\href {https://doi.org/10.1063/1.1388868} {\bibfield  {journal} {\bibinfo  {journal} {Applied Physics Letters}\ }\textbf {\bibinfo {volume} {79}},\ \bibinfo {pages} {705} (\bibinfo {year} {2001})}\BibitemShut {NoStop}%
\bibitem [{\citenamefont {Takesue}\ \emph {et~al.}(2007)\citenamefont {Takesue}, \citenamefont {Nam}, \citenamefont {Zhang}, \citenamefont {Hadfield}, \citenamefont {Honjo}, \citenamefont {Tamaki},\ and\ \citenamefont {Yamamoto}}]{Takesue2007}%
  \BibitemOpen
  \bibfield  {author} {\bibinfo {author} {\bibfnamefont {H.}~\bibnamefont {Takesue}}, \bibinfo {author} {\bibfnamefont {S.~W.}\ \bibnamefont {Nam}}, \bibinfo {author} {\bibfnamefont {Q.}~\bibnamefont {Zhang}}, \bibinfo {author} {\bibfnamefont {R.~H.}\ \bibnamefont {Hadfield}}, \bibinfo {author} {\bibfnamefont {T.}~\bibnamefont {Honjo}}, \bibinfo {author} {\bibfnamefont {K.}~\bibnamefont {Tamaki}},\ and\ \bibinfo {author} {\bibfnamefont {Y.}~\bibnamefont {Yamamoto}},\ }\bibfield  {title} {\bibinfo {title} {Quantum key distribution over a 40-{{dB}} channel loss using superconducting single-photon detectors},\ }\href {https://doi.org/10.1038/nphoton.2007.75} {\bibfield  {journal} {\bibinfo  {journal} {Nature Photonics}\ }\textbf {\bibinfo {volume} {1}},\ \bibinfo {pages} {343} (\bibinfo {year} {2007})}\BibitemShut {NoStop}%
\bibitem [{\citenamefont {Chen}\ \emph {et~al.}(2022)\citenamefont {Chen}, \citenamefont {Zhang}, \citenamefont {Liu}, \citenamefont {Jiang}, \citenamefont {Zhao}, \citenamefont {Zhang}, \citenamefont {Chen}, \citenamefont {Li}, \citenamefont {You}, \citenamefont {Wang}, \citenamefont {Chen}, \citenamefont {Wang}, \citenamefont {Zhang},\ and\ \citenamefont {Pan}}]{Chen2022}%
  \BibitemOpen
  \bibfield  {author} {\bibinfo {author} {\bibfnamefont {J.-P.}\ \bibnamefont {Chen}}, \bibinfo {author} {\bibfnamefont {C.}~\bibnamefont {Zhang}}, \bibinfo {author} {\bibfnamefont {Y.}~\bibnamefont {Liu}}, \bibinfo {author} {\bibfnamefont {C.}~\bibnamefont {Jiang}}, \bibinfo {author} {\bibfnamefont {D.-F.}\ \bibnamefont {Zhao}}, \bibinfo {author} {\bibfnamefont {W.-J.}\ \bibnamefont {Zhang}}, \bibinfo {author} {\bibfnamefont {F.-X.}\ \bibnamefont {Chen}}, \bibinfo {author} {\bibfnamefont {H.}~\bibnamefont {Li}}, \bibinfo {author} {\bibfnamefont {L.-X.}\ \bibnamefont {You}}, \bibinfo {author} {\bibfnamefont {Z.}~\bibnamefont {Wang}}, \bibinfo {author} {\bibfnamefont {Y.}~\bibnamefont {Chen}}, \bibinfo {author} {\bibfnamefont {X.-B.}\ \bibnamefont {Wang}}, \bibinfo {author} {\bibfnamefont {Q.}~\bibnamefont {Zhang}},\ and\ \bibinfo {author} {\bibfnamefont {J.-W.}\ \bibnamefont {Pan}},\ }\bibfield  {title} {\bibinfo {title} {Quantum {{Key Distribution}} over 658 km {{Fiber}} with {{Distributed Vibration
  Sensing}}},\ }\href {https://doi.org/10.1103/PhysRevLett.128.180502} {\bibfield  {journal} {\bibinfo  {journal} {Physical Review Letters}\ }\textbf {\bibinfo {volume} {128}},\ \bibinfo {pages} {180502} (\bibinfo {year} {2022})}\BibitemShut {NoStop}%
\bibitem [{\citenamefont {Liu}\ \emph {et~al.}(2023)\citenamefont {Liu}, \citenamefont {Zhang}, \citenamefont {Jiang}, \citenamefont {Chen}, \citenamefont {Zhang}, \citenamefont {Pan}, \citenamefont {Ma}, \citenamefont {Dong}, \citenamefont {Xiong}, \citenamefont {Zhang}, \citenamefont {Li}, \citenamefont {Wang}, \citenamefont {Wu}, \citenamefont {Chen}, \citenamefont {You}, \citenamefont {Wang}, \citenamefont {Zhang},\ and\ \citenamefont {Pan}}]{Liu2023}%
  \BibitemOpen
  \bibfield  {author} {\bibinfo {author} {\bibfnamefont {Y.}~\bibnamefont {Liu}}, \bibinfo {author} {\bibfnamefont {W.-J.}\ \bibnamefont {Zhang}}, \bibinfo {author} {\bibfnamefont {C.}~\bibnamefont {Jiang}}, \bibinfo {author} {\bibfnamefont {J.-P.}\ \bibnamefont {Chen}}, \bibinfo {author} {\bibfnamefont {C.}~\bibnamefont {Zhang}}, \bibinfo {author} {\bibfnamefont {W.-X.}\ \bibnamefont {Pan}}, \bibinfo {author} {\bibfnamefont {D.}~\bibnamefont {Ma}}, \bibinfo {author} {\bibfnamefont {H.}~\bibnamefont {Dong}}, \bibinfo {author} {\bibfnamefont {J.-M.}\ \bibnamefont {Xiong}}, \bibinfo {author} {\bibfnamefont {C.-J.}\ \bibnamefont {Zhang}}, \bibinfo {author} {\bibfnamefont {H.}~\bibnamefont {Li}}, \bibinfo {author} {\bibfnamefont {R.-C.}\ \bibnamefont {Wang}}, \bibinfo {author} {\bibfnamefont {J.}~\bibnamefont {Wu}}, \bibinfo {author} {\bibfnamefont {T.-Y.}\ \bibnamefont {Chen}}, \bibinfo {author} {\bibfnamefont {L.}~\bibnamefont {You}}, \bibinfo {author} {\bibfnamefont {X.-B.}\ \bibnamefont {Wang}}, \bibinfo
  {author} {\bibfnamefont {Q.}~\bibnamefont {Zhang}},\ and\ \bibinfo {author} {\bibfnamefont {J.-W.}\ \bibnamefont {Pan}},\ }\bibfield  {title} {\bibinfo {title} {Experimental {{Twin-Field Quantum Key Distribution}} over 1000 km {{Fiber Distance}}},\ }\href {https://doi.org/10.1103/PhysRevLett.130.210801} {\bibfield  {journal} {\bibinfo  {journal} {Physical Review Letters}\ }\textbf {\bibinfo {volume} {130}},\ \bibinfo {pages} {210801} (\bibinfo {year} {2023})}\BibitemShut {NoStop}%
\bibitem [{\citenamefont {Bussi{\`e}res}\ \emph {et~al.}(2014)\citenamefont {Bussi{\`e}res}, \citenamefont {Clausen}, \citenamefont {Tiranov}, \citenamefont {Korzh}, \citenamefont {Verma}, \citenamefont {Nam}, \citenamefont {Marsili}, \citenamefont {Ferrier}, \citenamefont {Goldner}, \citenamefont {Herrmann}, \citenamefont {Silberhorn}, \citenamefont {Sohler}, \citenamefont {Afzelius},\ and\ \citenamefont {Gisin}}]{Bussieres2014}%
  \BibitemOpen
  \bibfield  {author} {\bibinfo {author} {\bibfnamefont {F.}~\bibnamefont {Bussi{\`e}res}}, \bibinfo {author} {\bibfnamefont {C.}~\bibnamefont {Clausen}}, \bibinfo {author} {\bibfnamefont {A.}~\bibnamefont {Tiranov}}, \bibinfo {author} {\bibfnamefont {B.}~\bibnamefont {Korzh}}, \bibinfo {author} {\bibfnamefont {V.~B.}\ \bibnamefont {Verma}}, \bibinfo {author} {\bibfnamefont {S.~W.}\ \bibnamefont {Nam}}, \bibinfo {author} {\bibfnamefont {F.}~\bibnamefont {Marsili}}, \bibinfo {author} {\bibfnamefont {A.}~\bibnamefont {Ferrier}}, \bibinfo {author} {\bibfnamefont {P.}~\bibnamefont {Goldner}}, \bibinfo {author} {\bibfnamefont {H.}~\bibnamefont {Herrmann}}, \bibinfo {author} {\bibfnamefont {C.}~\bibnamefont {Silberhorn}}, \bibinfo {author} {\bibfnamefont {W.}~\bibnamefont {Sohler}}, \bibinfo {author} {\bibfnamefont {M.}~\bibnamefont {Afzelius}},\ and\ \bibinfo {author} {\bibfnamefont {N.}~\bibnamefont {Gisin}},\ }\bibfield  {title} {\bibinfo {title} {Quantum teleportation from a telecom-wavelength photon to a
  solid-state quantum memory},\ }\href {https://doi.org/10.1038/nphoton.2014.215} {\bibfield  {journal} {\bibinfo  {journal} {Nature Photonics}\ }\textbf {\bibinfo {volume} {8}},\ \bibinfo {pages} {775} (\bibinfo {year} {2014})}\BibitemShut {NoStop}%
\bibitem [{\citenamefont {Takesue}\ \emph {et~al.}(2015)\citenamefont {Takesue}, \citenamefont {Dyer}, \citenamefont {Stevens}, \citenamefont {Verma}, \citenamefont {Mirin},\ and\ \citenamefont {Nam}}]{Takesue2015}%
  \BibitemOpen
  \bibfield  {author} {\bibinfo {author} {\bibfnamefont {H.}~\bibnamefont {Takesue}}, \bibinfo {author} {\bibfnamefont {S.~D.}\ \bibnamefont {Dyer}}, \bibinfo {author} {\bibfnamefont {M.~J.}\ \bibnamefont {Stevens}}, \bibinfo {author} {\bibfnamefont {V.}~\bibnamefont {Verma}}, \bibinfo {author} {\bibfnamefont {R.~P.}\ \bibnamefont {Mirin}},\ and\ \bibinfo {author} {\bibfnamefont {S.~W.}\ \bibnamefont {Nam}},\ }\bibfield  {title} {\bibinfo {title} {Quantum teleportation over 100 km of fiber using highly efficient superconducting nanowire single-photon detectors},\ }\href {https://doi.org/10.1364/OPTICA.2.000832} {\bibfield  {journal} {\bibinfo  {journal} {Optica}\ }\textbf {\bibinfo {volume} {2}},\ \bibinfo {pages} {832} (\bibinfo {year} {2015})}\BibitemShut {NoStop}%
\bibitem [{\citenamefont {Shibata}\ \emph {et~al.}(2014)\citenamefont {Shibata}, \citenamefont {Honjo},\ and\ \citenamefont {Shimizu}}]{Shibata2014}%
  \BibitemOpen
  \bibfield  {author} {\bibinfo {author} {\bibfnamefont {H.}~\bibnamefont {Shibata}}, \bibinfo {author} {\bibfnamefont {T.}~\bibnamefont {Honjo}},\ and\ \bibinfo {author} {\bibfnamefont {K.}~\bibnamefont {Shimizu}},\ }\bibfield  {title} {\bibinfo {title} {Quantum key distribution over a 72 {{dB}} channel loss using ultralow dark count superconducting single-photon detectors},\ }\href {https://doi.org/10.1364/OL.39.005078} {\bibfield  {journal} {\bibinfo  {journal} {Optics Letters}\ }\textbf {\bibinfo {volume} {39}},\ \bibinfo {pages} {5078} (\bibinfo {year} {2014})}\BibitemShut {NoStop}%
\bibitem [{\citenamefont {Valivarthi}\ \emph {et~al.}(2016)\citenamefont {Valivarthi}, \citenamefont {Puigibert}, \citenamefont {Zhou}, \citenamefont {Aguilar}, \citenamefont {Verma}, \citenamefont {Marsili}, \citenamefont {Shaw}, \citenamefont {Nam}, \citenamefont {Oblak},\ and\ \citenamefont {Tittel}}]{Valivarthi2016}%
  \BibitemOpen
  \bibfield  {author} {\bibinfo {author} {\bibfnamefont {R.}~\bibnamefont {Valivarthi}}, \bibinfo {author} {\bibfnamefont {M.~G.}\ \bibnamefont {Puigibert}}, \bibinfo {author} {\bibfnamefont {Q.}~\bibnamefont {Zhou}}, \bibinfo {author} {\bibfnamefont {G.~H.}\ \bibnamefont {Aguilar}}, \bibinfo {author} {\bibfnamefont {V.~B.}\ \bibnamefont {Verma}}, \bibinfo {author} {\bibfnamefont {F.}~\bibnamefont {Marsili}}, \bibinfo {author} {\bibfnamefont {M.~D.}\ \bibnamefont {Shaw}}, \bibinfo {author} {\bibfnamefont {S.~W.}\ \bibnamefont {Nam}}, \bibinfo {author} {\bibfnamefont {D.}~\bibnamefont {Oblak}},\ and\ \bibinfo {author} {\bibfnamefont {W.}~\bibnamefont {Tittel}},\ }\bibfield  {title} {\bibinfo {title} {Quantum teleportation across a metropolitan fibre network},\ }\href {https://doi.org/10.1038/nphoton.2016.180} {\bibfield  {journal} {\bibinfo  {journal} {Nature Photonics}\ }\textbf {\bibinfo {volume} {10}},\ \bibinfo {pages} {676} (\bibinfo {year} {2016})}\BibitemShut {NoStop}%
\bibitem [{\citenamefont {Natarajan}\ \emph {et~al.}(2012)\citenamefont {Natarajan}, \citenamefont {Tanner},\ and\ \citenamefont {Hadfield}}]{Natarajan2012}%
  \BibitemOpen
  \bibfield  {author} {\bibinfo {author} {\bibfnamefont {C.~M.}\ \bibnamefont {Natarajan}}, \bibinfo {author} {\bibfnamefont {M.~G.}\ \bibnamefont {Tanner}},\ and\ \bibinfo {author} {\bibfnamefont {R.~H.}\ \bibnamefont {Hadfield}},\ }\bibfield  {title} {\bibinfo {title} {Superconducting nanowire single-photon detectors: Physics and applications},\ }\href {https://doi.org/10.1088/0953-2048/25/6/063001} {\bibfield  {journal} {\bibinfo  {journal} {Superconductor Science and Technology}\ }\textbf {\bibinfo {volume} {25}},\ \bibinfo {pages} {063001} (\bibinfo {year} {2012})}\BibitemShut {NoStop}%
\bibitem [{\citenamefont {Wollman}\ \emph {et~al.}(2021)\citenamefont {Wollman}, \citenamefont {Verma}, \citenamefont {Walter}, \citenamefont {Chiles}, \citenamefont {Korzh}, \citenamefont {Allmaras}, \citenamefont {Zhai}, \citenamefont {Lita}, \citenamefont {McCaughan}, \citenamefont {Schmidt}, \citenamefont {Frasca}, \citenamefont {Mirin}, \citenamefont {Nam},\ and\ \citenamefont {Shaw}}]{Wollman2021}%
  \BibitemOpen
  \bibfield  {author} {\bibinfo {author} {\bibfnamefont {E.~E.}\ \bibnamefont {Wollman}}, \bibinfo {author} {\bibfnamefont {V.~B.}\ \bibnamefont {Verma}}, \bibinfo {author} {\bibfnamefont {A.~B.}\ \bibnamefont {Walter}}, \bibinfo {author} {\bibfnamefont {J.}~\bibnamefont {Chiles}}, \bibinfo {author} {\bibfnamefont {B.}~\bibnamefont {Korzh}}, \bibinfo {author} {\bibfnamefont {J.~P.}\ \bibnamefont {Allmaras}}, \bibinfo {author} {\bibfnamefont {Y.}~\bibnamefont {Zhai}}, \bibinfo {author} {\bibfnamefont {A.~E.}\ \bibnamefont {Lita}}, \bibinfo {author} {\bibfnamefont {A.~N.}\ \bibnamefont {McCaughan}}, \bibinfo {author} {\bibfnamefont {E.}~\bibnamefont {Schmidt}}, \bibinfo {author} {\bibfnamefont {S.}~\bibnamefont {Frasca}}, \bibinfo {author} {\bibfnamefont {R.~P.}\ \bibnamefont {Mirin}}, \bibinfo {author} {\bibfnamefont {S.~W.}\ \bibnamefont {Nam}},\ and\ \bibinfo {author} {\bibfnamefont {M.~D.}\ \bibnamefont {Shaw}},\ }\bibfield  {title} {\bibinfo {title} {Recent advances in superconducting nanowire
  single-photon detector technology for exoplanet transit spectroscopy in the mid-infrared},\ }\href {https://doi.org/10.1117/1.JATIS.7.1.011004} {\bibfield  {journal} {\bibinfo  {journal} {Journal of Astronomical Telescopes, Instruments, and Systems}\ }\textbf {\bibinfo {volume} {7}},\ \bibinfo {pages} {011004} (\bibinfo {year} {2021})}\BibitemShut {NoStop}%
\bibitem [{\citenamefont {Grein}\ \emph {et~al.}(2015)\citenamefont {Grein}, \citenamefont {Kerman}, \citenamefont {Dauler}, \citenamefont {Willis}, \citenamefont {Romkey}, \citenamefont {Molnar}, \citenamefont {Robinson}, \citenamefont {Murphy},\ and\ \citenamefont {Boroson}}]{Grein2015}%
  \BibitemOpen
  \bibfield  {author} {\bibinfo {author} {\bibfnamefont {M.~E.}\ \bibnamefont {Grein}}, \bibinfo {author} {\bibfnamefont {A.~J.}\ \bibnamefont {Kerman}}, \bibinfo {author} {\bibfnamefont {E.~A.}\ \bibnamefont {Dauler}}, \bibinfo {author} {\bibfnamefont {M.~M.}\ \bibnamefont {Willis}}, \bibinfo {author} {\bibfnamefont {B.}~\bibnamefont {Romkey}}, \bibinfo {author} {\bibfnamefont {R.~J.}\ \bibnamefont {Molnar}}, \bibinfo {author} {\bibfnamefont {B.~S.}\ \bibnamefont {Robinson}}, \bibinfo {author} {\bibfnamefont {D.~V.}\ \bibnamefont {Murphy}},\ and\ \bibinfo {author} {\bibfnamefont {D.~M.}\ \bibnamefont {Boroson}},\ }\bibfield  {title} {\bibinfo {title} {An optical receiver for the {{Lunar Laser Communication Demonstration}} based on photon-counting superconducting nanowires},\ }\href {https://doi.org/10.1117/12.2179781} {\bibfield  {journal} {\bibinfo  {journal} {Advanced Photon Counting Techniques IX}\ }\textbf {\bibinfo {volume} {9492}},\ \bibinfo {pages} {949208} (\bibinfo {year} {2015})}\BibitemShut
  {NoStop}%
\bibitem [{\citenamefont {Biswas}\ \emph {et~al.}(2017)\citenamefont {Biswas}, \citenamefont {Srinivasan}, \citenamefont {Rogalin}, \citenamefont {Piazzolla}, \citenamefont {Liu}, \citenamefont {Schratz}, \citenamefont {Wong}, \citenamefont {Alerstam}, \citenamefont {Wright}, \citenamefont {Roberts}, \citenamefont {Kovalik}, \citenamefont {Ortiz}, \citenamefont {{Na-Nakornpanom}}, \citenamefont {Shaw}, \citenamefont {Okino}, \citenamefont {Andrews}, \citenamefont {Peng}, \citenamefont {Orozco},\ and\ \citenamefont {Klipstein}}]{Biswas2017}%
  \BibitemOpen
  \bibfield  {author} {\bibinfo {author} {\bibfnamefont {A.}~\bibnamefont {Biswas}}, \bibinfo {author} {\bibfnamefont {M.}~\bibnamefont {Srinivasan}}, \bibinfo {author} {\bibfnamefont {R.}~\bibnamefont {Rogalin}}, \bibinfo {author} {\bibfnamefont {S.}~\bibnamefont {Piazzolla}}, \bibinfo {author} {\bibfnamefont {J.}~\bibnamefont {Liu}}, \bibinfo {author} {\bibfnamefont {B.}~\bibnamefont {Schratz}}, \bibinfo {author} {\bibfnamefont {A.}~\bibnamefont {Wong}}, \bibinfo {author} {\bibfnamefont {E.}~\bibnamefont {Alerstam}}, \bibinfo {author} {\bibfnamefont {M.}~\bibnamefont {Wright}}, \bibinfo {author} {\bibfnamefont {W.~T.}\ \bibnamefont {Roberts}}, \bibinfo {author} {\bibfnamefont {J.}~\bibnamefont {Kovalik}}, \bibinfo {author} {\bibfnamefont {G.}~\bibnamefont {Ortiz}}, \bibinfo {author} {\bibfnamefont {A.}~\bibnamefont {{Na-Nakornpanom}}}, \bibinfo {author} {\bibfnamefont {M.}~\bibnamefont {Shaw}}, \bibinfo {author} {\bibfnamefont {C.}~\bibnamefont {Okino}}, \bibinfo {author} {\bibfnamefont {K.}~\bibnamefont
  {Andrews}}, \bibinfo {author} {\bibfnamefont {M.}~\bibnamefont {Peng}}, \bibinfo {author} {\bibfnamefont {D.}~\bibnamefont {Orozco}},\ and\ \bibinfo {author} {\bibfnamefont {W.}~\bibnamefont {Klipstein}},\ }\bibfield  {title} {\bibinfo {title} {Status of {{NASA}}'s deep space optical communication technology demonstration},\ }in\ \href {https://doi.org/10.1109/ICSOS.2017.8357206} {\emph {\bibinfo {booktitle} {2017 {{IEEE International Conference}} on {{Space Optical Systems}} and {{Applications}} ({{ICSOS}})}}}\ (\bibinfo  {publisher} {IEEE},\ \bibinfo {address} {Naha},\ \bibinfo {year} {2017})\ pp.\ \bibinfo {pages} {23--27}\BibitemShut {NoStop}%
\bibitem [{\citenamefont {Xia}\ \emph {et~al.}(2021)\citenamefont {Xia}, \citenamefont {Gevers}, \citenamefont {Fognini}, \citenamefont {Mok}, \citenamefont {Li}, \citenamefont {Akbari}, \citenamefont {Zadeh}, \citenamefont {{Qin-Dregely}},\ and\ \citenamefont {Xu}}]{Xia2021}%
  \BibitemOpen
  \bibfield  {author} {\bibinfo {author} {\bibfnamefont {F.}~\bibnamefont {Xia}}, \bibinfo {author} {\bibfnamefont {M.}~\bibnamefont {Gevers}}, \bibinfo {author} {\bibfnamefont {A.}~\bibnamefont {Fognini}}, \bibinfo {author} {\bibfnamefont {A.~T.}\ \bibnamefont {Mok}}, \bibinfo {author} {\bibfnamefont {B.}~\bibnamefont {Li}}, \bibinfo {author} {\bibfnamefont {N.}~\bibnamefont {Akbari}}, \bibinfo {author} {\bibfnamefont {I.~E.}\ \bibnamefont {Zadeh}}, \bibinfo {author} {\bibfnamefont {J.}~\bibnamefont {{Qin-Dregely}}},\ and\ \bibinfo {author} {\bibfnamefont {C.}~\bibnamefont {Xu}},\ }\bibfield  {title} {\bibinfo {title} {Short-{{Wave Infrared Confocal Fluorescence Imaging}} of {{Deep Mouse Brain}} with a {{Superconducting Nanowire Single-Photon Detector}}},\ }\href {https://doi.org/10.1021/acsphotonics.1c01018} {\bibfield  {journal} {\bibinfo  {journal} {ACS Photonics}\ }\textbf {\bibinfo {volume} {8}},\ \bibinfo {pages} {2800} (\bibinfo {year} {2021})}\BibitemShut {NoStop}%
\bibitem [{\citenamefont {Tamimi}\ \emph {et~al.}(2023)\citenamefont {Tamimi}, \citenamefont {Caldarola}, \citenamefont {Hambura}, \citenamefont {Boffi}, \citenamefont {Noordzij}, \citenamefont {Los}, \citenamefont {Guardiani}, \citenamefont {Kooiman}, \citenamefont {Wang}, \citenamefont {Kieser}, \citenamefont {Braun}, \citenamefont {Fognini},\ and\ \citenamefont {Prevedel}}]{Tamimi2023}%
  \BibitemOpen
  \bibfield  {author} {\bibinfo {author} {\bibfnamefont {A.}~\bibnamefont {Tamimi}}, \bibinfo {author} {\bibfnamefont {M.}~\bibnamefont {Caldarola}}, \bibinfo {author} {\bibfnamefont {S.}~\bibnamefont {Hambura}}, \bibinfo {author} {\bibfnamefont {J.~C.}\ \bibnamefont {Boffi}}, \bibinfo {author} {\bibfnamefont {N.}~\bibnamefont {Noordzij}}, \bibinfo {author} {\bibfnamefont {J.~W.~N.}\ \bibnamefont {Los}}, \bibinfo {author} {\bibfnamefont {A.}~\bibnamefont {Guardiani}}, \bibinfo {author} {\bibfnamefont {H.}~\bibnamefont {Kooiman}}, \bibinfo {author} {\bibfnamefont {L.}~\bibnamefont {Wang}}, \bibinfo {author} {\bibfnamefont {C.}~\bibnamefont {Kieser}}, \bibinfo {author} {\bibfnamefont {F.}~\bibnamefont {Braun}}, \bibinfo {author} {\bibfnamefont {A.}~\bibnamefont {Fognini}},\ and\ \bibinfo {author} {\bibfnamefont {R.}~\bibnamefont {Prevedel}},\ }\href@noop {} {\bibinfo {title} {Deep mouse brain two-photon near-infrared fluorescence imaging using a superconducting nanowire single-photon detector array}} (\bibinfo
  {year} {2023})\BibitemShut {NoStop}%
\bibitem [{\citenamefont {Ozana}\ \emph {et~al.}(2021)\citenamefont {Ozana}, \citenamefont {Zavriyev}, \citenamefont {Mazumder}, \citenamefont {Robinson}, \citenamefont {Kaya}, \citenamefont {Blackwell}, \citenamefont {Carp},\ and\ \citenamefont {Franceschini}}]{Ozana2021}%
  \BibitemOpen
  \bibfield  {author} {\bibinfo {author} {\bibfnamefont {N.}~\bibnamefont {Ozana}}, \bibinfo {author} {\bibfnamefont {A.~I.}\ \bibnamefont {Zavriyev}}, \bibinfo {author} {\bibfnamefont {D.}~\bibnamefont {Mazumder}}, \bibinfo {author} {\bibfnamefont {M.~B.}\ \bibnamefont {Robinson}}, \bibinfo {author} {\bibfnamefont {K.}~\bibnamefont {Kaya}}, \bibinfo {author} {\bibfnamefont {M.~H.}\ \bibnamefont {Blackwell}}, \bibinfo {author} {\bibfnamefont {S.~A.}\ \bibnamefont {Carp}},\ and\ \bibinfo {author} {\bibfnamefont {M.~A.}\ \bibnamefont {Franceschini}},\ }\bibfield  {title} {\bibinfo {title} {Superconducting nanowire single-photon sensing of cerebral blood flow},\ }\href {https://doi.org/10.1117/1.NPh.8.3.035006} {\bibfield  {journal} {\bibinfo  {journal} {Neurophotonics}\ }\textbf {\bibinfo {volume} {8}},\ \bibinfo {pages} {035006} (\bibinfo {year} {2021})}\BibitemShut {NoStop}%
\bibitem [{\citenamefont {Marsili}\ \emph {et~al.}(2012)\citenamefont {Marsili}, \citenamefont {Bellei}, \citenamefont {Najafi}, \citenamefont {Dane}, \citenamefont {Dauler}, \citenamefont {Molnar},\ and\ \citenamefont {Berggren}}]{Marsili2012a}%
  \BibitemOpen
  \bibfield  {author} {\bibinfo {author} {\bibfnamefont {F.}~\bibnamefont {Marsili}}, \bibinfo {author} {\bibfnamefont {F.}~\bibnamefont {Bellei}}, \bibinfo {author} {\bibfnamefont {F.}~\bibnamefont {Najafi}}, \bibinfo {author} {\bibfnamefont {A.~E.}\ \bibnamefont {Dane}}, \bibinfo {author} {\bibfnamefont {E.~A.}\ \bibnamefont {Dauler}}, \bibinfo {author} {\bibfnamefont {R.~J.}\ \bibnamefont {Molnar}},\ and\ \bibinfo {author} {\bibfnamefont {K.~K.}\ \bibnamefont {Berggren}},\ }\bibfield  {title} {\bibinfo {title} {Efficient {{Single Photon Detection}} from 500nm to 5{\textmu}m {{Wavelength}}},\ }\href {https://doi.org/10.1021/nl302245n} {\bibfield  {journal} {\bibinfo  {journal} {Nano Letters}\ }\textbf {\bibinfo {volume} {12}},\ \bibinfo {pages} {4799} (\bibinfo {year} {2012})}\BibitemShut {NoStop}%
\bibitem [{\citenamefont {Korneev}\ \emph {et~al.}(2012)\citenamefont {Korneev}, \citenamefont {Korneeva}, \citenamefont {Florya}, \citenamefont {Voronov},\ and\ \citenamefont {Goltsman}}]{Korneev2012}%
  \BibitemOpen
  \bibfield  {author} {\bibinfo {author} {\bibfnamefont {A.}~\bibnamefont {Korneev}}, \bibinfo {author} {\bibfnamefont {{\relax Yu}.}~\bibnamefont {Korneeva}}, \bibinfo {author} {\bibfnamefont {I.}~\bibnamefont {Florya}}, \bibinfo {author} {\bibfnamefont {B.}~\bibnamefont {Voronov}},\ and\ \bibinfo {author} {\bibfnamefont {G.}~\bibnamefont {Goltsman}},\ }\bibfield  {title} {\bibinfo {title} {{{NbN Nanowire Superconducting Single-Photon Detector}} for {{Mid-Infrared}}},\ }\href {https://doi.org/10.1016/j.phpro.2012.06.215} {\bibfield  {journal} {\bibinfo  {journal} {Physics Procedia}\ }\textbf {\bibinfo {volume} {36}},\ \bibinfo {pages} {72} (\bibinfo {year} {2012})}\BibitemShut {NoStop}%
\bibitem [{\citenamefont {Shibata}\ \emph {et~al.}(2015)\citenamefont {Shibata}, \citenamefont {Shimizu}, \citenamefont {Takesue},\ and\ \citenamefont {Tokura}}]{Shibata2015}%
  \BibitemOpen
  \bibfield  {author} {\bibinfo {author} {\bibfnamefont {H.}~\bibnamefont {Shibata}}, \bibinfo {author} {\bibfnamefont {K.}~\bibnamefont {Shimizu}}, \bibinfo {author} {\bibfnamefont {H.}~\bibnamefont {Takesue}},\ and\ \bibinfo {author} {\bibfnamefont {Y.}~\bibnamefont {Tokura}},\ }\bibfield  {title} {\bibinfo {title} {Ultimate low system dark-count rate for superconducting nanowire single-photon detector},\ }\href {https://doi.org/10.1364/ol.40.003428} {\bibfield  {journal} {\bibinfo  {journal} {Optics Letters}\ }\textbf {\bibinfo {volume} {40}},\ \bibinfo {pages} {3428} (\bibinfo {year} {2015})}\BibitemShut {NoStop}%
\bibitem [{\citenamefont {Korzh}\ \emph {et~al.}(2020)\citenamefont {Korzh}, \citenamefont {Zhao}, \citenamefont {Allmaras}, \citenamefont {Frasca}, \citenamefont {Autry}, \citenamefont {Bersin}, \citenamefont {Beyer}, \citenamefont {Briggs}, \citenamefont {Bumble}, \citenamefont {Colangelo}, \citenamefont {Crouch}, \citenamefont {Dane}, \citenamefont {Gerrits}, \citenamefont {Lita}, \citenamefont {Marsili}, \citenamefont {Moody}, \citenamefont {Pe{\~n}a}, \citenamefont {Ramirez}, \citenamefont {Rezac}, \citenamefont {Sinclair}, \citenamefont {Stevens}, \citenamefont {Velasco}, \citenamefont {Verma}, \citenamefont {Wollman}, \citenamefont {Xie}, \citenamefont {Zhu}, \citenamefont {Hale}, \citenamefont {Spiropulu}, \citenamefont {Silverman}, \citenamefont {Mirin}, \citenamefont {Nam}, \citenamefont {Kozorezov}, \citenamefont {Shaw},\ and\ \citenamefont {Berggren}}]{Korzh2020}%
  \BibitemOpen
  \bibfield  {author} {\bibinfo {author} {\bibfnamefont {B.}~\bibnamefont {Korzh}}, \bibinfo {author} {\bibfnamefont {Q.~Y.}\ \bibnamefont {Zhao}}, \bibinfo {author} {\bibfnamefont {J.~P.}\ \bibnamefont {Allmaras}}, \bibinfo {author} {\bibfnamefont {S.}~\bibnamefont {Frasca}}, \bibinfo {author} {\bibfnamefont {T.~M.}\ \bibnamefont {Autry}}, \bibinfo {author} {\bibfnamefont {E.~A.}\ \bibnamefont {Bersin}}, \bibinfo {author} {\bibfnamefont {A.~D.}\ \bibnamefont {Beyer}}, \bibinfo {author} {\bibfnamefont {R.~M.}\ \bibnamefont {Briggs}}, \bibinfo {author} {\bibfnamefont {B.}~\bibnamefont {Bumble}}, \bibinfo {author} {\bibfnamefont {M.}~\bibnamefont {Colangelo}}, \bibinfo {author} {\bibfnamefont {G.~M.}\ \bibnamefont {Crouch}}, \bibinfo {author} {\bibfnamefont {A.~E.}\ \bibnamefont {Dane}}, \bibinfo {author} {\bibfnamefont {T.}~\bibnamefont {Gerrits}}, \bibinfo {author} {\bibfnamefont {A.~E.}\ \bibnamefont {Lita}}, \bibinfo {author} {\bibfnamefont {F.}~\bibnamefont {Marsili}}, \bibinfo {author} {\bibfnamefont
  {G.}~\bibnamefont {Moody}}, \bibinfo {author} {\bibfnamefont {C.}~\bibnamefont {Pe{\~n}a}}, \bibinfo {author} {\bibfnamefont {E.}~\bibnamefont {Ramirez}}, \bibinfo {author} {\bibfnamefont {J.~D.}\ \bibnamefont {Rezac}}, \bibinfo {author} {\bibfnamefont {N.}~\bibnamefont {Sinclair}}, \bibinfo {author} {\bibfnamefont {M.~J.}\ \bibnamefont {Stevens}}, \bibinfo {author} {\bibfnamefont {A.~E.}\ \bibnamefont {Velasco}}, \bibinfo {author} {\bibfnamefont {V.~B.}\ \bibnamefont {Verma}}, \bibinfo {author} {\bibfnamefont {E.~E.}\ \bibnamefont {Wollman}}, \bibinfo {author} {\bibfnamefont {S.}~\bibnamefont {Xie}}, \bibinfo {author} {\bibfnamefont {D.}~\bibnamefont {Zhu}}, \bibinfo {author} {\bibfnamefont {P.~D.}\ \bibnamefont {Hale}}, \bibinfo {author} {\bibfnamefont {M.}~\bibnamefont {Spiropulu}}, \bibinfo {author} {\bibfnamefont {K.~L.}\ \bibnamefont {Silverman}}, \bibinfo {author} {\bibfnamefont {R.~P.}\ \bibnamefont {Mirin}}, \bibinfo {author} {\bibfnamefont {S.~W.}\ \bibnamefont {Nam}}, \bibinfo {author}
  {\bibfnamefont {A.~G.}\ \bibnamefont {Kozorezov}}, \bibinfo {author} {\bibfnamefont {M.~D.}\ \bibnamefont {Shaw}},\ and\ \bibinfo {author} {\bibfnamefont {K.~K.}\ \bibnamefont {Berggren}},\ }\bibfield  {title} {\bibinfo {title} {Demonstration of sub-3 ps temporal resolution with a superconducting nanowire single-photon detector},\ }\href {https://doi.org/10.1038/s41566-020-0589-x} {\bibfield  {journal} {\bibinfo  {journal} {Nature Photonics}\ }\textbf {\bibinfo {volume} {14}},\ \bibinfo {pages} {250} (\bibinfo {year} {2020})}\BibitemShut {NoStop}%
\bibitem [{\citenamefont {Cherednichenko}\ \emph {et~al.}(2021)\citenamefont {Cherednichenko}, \citenamefont {Acharya}, \citenamefont {Novoselov},\ and\ \citenamefont {Drakinskiy}}]{Cherednichenko2021}%
  \BibitemOpen
  \bibfield  {author} {\bibinfo {author} {\bibfnamefont {S.}~\bibnamefont {Cherednichenko}}, \bibinfo {author} {\bibfnamefont {N.}~\bibnamefont {Acharya}}, \bibinfo {author} {\bibfnamefont {E.}~\bibnamefont {Novoselov}},\ and\ \bibinfo {author} {\bibfnamefont {V.}~\bibnamefont {Drakinskiy}},\ }\bibfield  {title} {\bibinfo {title} {Low kinetic inductance superconducting {{MgB}} {\textsubscript{2}} nanowires with a 130 ps relaxation time for single-photon detection applications},\ }\href {https://doi.org/10.1088/1361-6668/abdeda} {\bibfield  {journal} {\bibinfo  {journal} {Superconductor Science and Technology}\ }\textbf {\bibinfo {volume} {34}},\ \bibinfo {pages} {044001} (\bibinfo {year} {2021})}\BibitemShut {NoStop}%
\bibitem [{\citenamefont {Slussarenko}\ and\ \citenamefont {Pryde}(2019)}]{Slussarenko2019}%
  \BibitemOpen
  \bibfield  {author} {\bibinfo {author} {\bibfnamefont {S.}~\bibnamefont {Slussarenko}}\ and\ \bibinfo {author} {\bibfnamefont {G.~J.}\ \bibnamefont {Pryde}},\ }\bibfield  {title} {\bibinfo {title} {Photonic quantum information processing: {{A}} concise review},\ }\href {https://doi.org/10.1063/1.5115814} {\bibfield  {journal} {\bibinfo  {journal} {Applied Physics Reviews}\ }\textbf {\bibinfo {volume} {6}},\ \bibinfo {pages} {041303} (\bibinfo {year} {2019})}\BibitemShut {NoStop}%
\bibitem [{\citenamefont {Gyger}\ \emph {et~al.}(2021)\citenamefont {Gyger}, \citenamefont {Zichi}, \citenamefont {Schweickert}, \citenamefont {Elshaari}, \citenamefont {Steinhauer}, \citenamefont {Covre Da~Silva}, \citenamefont {Rastelli}, \citenamefont {Zwiller}, \citenamefont {J{\"o}ns},\ and\ \citenamefont {{Errando-Herranz}}}]{Gyger2021}%
  \BibitemOpen
  \bibfield  {author} {\bibinfo {author} {\bibfnamefont {S.}~\bibnamefont {Gyger}}, \bibinfo {author} {\bibfnamefont {J.}~\bibnamefont {Zichi}}, \bibinfo {author} {\bibfnamefont {L.}~\bibnamefont {Schweickert}}, \bibinfo {author} {\bibfnamefont {A.~W.}\ \bibnamefont {Elshaari}}, \bibinfo {author} {\bibfnamefont {S.}~\bibnamefont {Steinhauer}}, \bibinfo {author} {\bibfnamefont {S.~F.}\ \bibnamefont {Covre Da~Silva}}, \bibinfo {author} {\bibfnamefont {A.}~\bibnamefont {Rastelli}}, \bibinfo {author} {\bibfnamefont {V.}~\bibnamefont {Zwiller}}, \bibinfo {author} {\bibfnamefont {K.~D.}\ \bibnamefont {J{\"o}ns}},\ and\ \bibinfo {author} {\bibfnamefont {C.}~\bibnamefont {{Errando-Herranz}}},\ }\bibfield  {title} {\bibinfo {title} {Reconfigurable photonics with on-chip single-photon detectors},\ }\href {https://doi.org/10.1038/s41467-021-21624-3} {\bibfield  {journal} {\bibinfo  {journal} {Nature Communications}\ }\textbf {\bibinfo {volume} {12}},\ \bibinfo {pages} {1408} (\bibinfo {year} {2021})}\BibitemShut
  {NoStop}%
\bibitem [{\citenamefont {Ferrari}\ \emph {et~al.}(2018)\citenamefont {Ferrari}, \citenamefont {Schuck},\ and\ \citenamefont {Pernice}}]{Ferrari2018}%
  \BibitemOpen
  \bibfield  {author} {\bibinfo {author} {\bibfnamefont {S.}~\bibnamefont {Ferrari}}, \bibinfo {author} {\bibfnamefont {C.}~\bibnamefont {Schuck}},\ and\ \bibinfo {author} {\bibfnamefont {W.}~\bibnamefont {Pernice}},\ }\bibfield  {title} {\bibinfo {title} {Waveguide-integrated superconducting nanowire single-photon detectors},\ }\href {https://doi.org/10.1515/nanoph-2018-0059} {\bibfield  {journal} {\bibinfo  {journal} {Nanophotonics}\ }\textbf {\bibinfo {volume} {7}},\ \bibinfo {pages} {1725} (\bibinfo {year} {2018})}\BibitemShut {NoStop}%
\bibitem [{\citenamefont {Sprengers}\ \emph {et~al.}(2011)\citenamefont {Sprengers}, \citenamefont {Gaggero}, \citenamefont {Sahin}, \citenamefont {Jahanmirinejad}, \citenamefont {Frucci}, \citenamefont {Mattioli}, \citenamefont {Leoni}, \citenamefont {Beetz}, \citenamefont {Lermer}, \citenamefont {Kamp}, \citenamefont {H{\"o}fling}, \citenamefont {Sanjines},\ and\ \citenamefont {Fiore}}]{Sprengers2011}%
  \BibitemOpen
  \bibfield  {author} {\bibinfo {author} {\bibfnamefont {J.~P.}\ \bibnamefont {Sprengers}}, \bibinfo {author} {\bibfnamefont {A.}~\bibnamefont {Gaggero}}, \bibinfo {author} {\bibfnamefont {D.}~\bibnamefont {Sahin}}, \bibinfo {author} {\bibfnamefont {S.}~\bibnamefont {Jahanmirinejad}}, \bibinfo {author} {\bibfnamefont {G.}~\bibnamefont {Frucci}}, \bibinfo {author} {\bibfnamefont {F.}~\bibnamefont {Mattioli}}, \bibinfo {author} {\bibfnamefont {R.}~\bibnamefont {Leoni}}, \bibinfo {author} {\bibfnamefont {J.}~\bibnamefont {Beetz}}, \bibinfo {author} {\bibfnamefont {M.}~\bibnamefont {Lermer}}, \bibinfo {author} {\bibfnamefont {M.}~\bibnamefont {Kamp}}, \bibinfo {author} {\bibfnamefont {S.}~\bibnamefont {H{\"o}fling}}, \bibinfo {author} {\bibfnamefont {R.}~\bibnamefont {Sanjines}},\ and\ \bibinfo {author} {\bibfnamefont {A.}~\bibnamefont {Fiore}},\ }\bibfield  {title} {\bibinfo {title} {Waveguide superconducting single-photon detectors for integrated quantum photonic circuits},\ }\href
  {https://doi.org/10.1063/1.3657518} {\bibfield  {journal} {\bibinfo  {journal} {Applied Physics Letters}\ }\textbf {\bibinfo {volume} {99}},\ \bibinfo {pages} {181110} (\bibinfo {year} {2011})}\BibitemShut {NoStop}%
\bibitem [{\citenamefont {Reithmaier}\ \emph {et~al.}(2013)\citenamefont {Reithmaier}, \citenamefont {Lichtmannecker}, \citenamefont {Reichert}, \citenamefont {Hasch}, \citenamefont {M{\"u}ller}, \citenamefont {Bichler}, \citenamefont {Gross},\ and\ \citenamefont {Finley}}]{Reithmaier2013}%
  \BibitemOpen
  \bibfield  {author} {\bibinfo {author} {\bibfnamefont {G.}~\bibnamefont {Reithmaier}}, \bibinfo {author} {\bibfnamefont {S.}~\bibnamefont {Lichtmannecker}}, \bibinfo {author} {\bibfnamefont {T.}~\bibnamefont {Reichert}}, \bibinfo {author} {\bibfnamefont {P.}~\bibnamefont {Hasch}}, \bibinfo {author} {\bibfnamefont {K.}~\bibnamefont {M{\"u}ller}}, \bibinfo {author} {\bibfnamefont {M.}~\bibnamefont {Bichler}}, \bibinfo {author} {\bibfnamefont {R.}~\bibnamefont {Gross}},\ and\ \bibinfo {author} {\bibfnamefont {J.~J.}\ \bibnamefont {Finley}},\ }\bibfield  {title} {\bibinfo {title} {On-chip time resolved detection of quantum dot emission using integrated superconducting single photon detectors},\ }\href {https://doi.org/10.1038/srep01901} {\bibfield  {journal} {\bibinfo  {journal} {Scientific Reports}\ }\textbf {\bibinfo {volume} {3}},\ \bibinfo {pages} {1901} (\bibinfo {year} {2013})}\BibitemShut {NoStop}%
\bibitem [{\citenamefont {Reithmaier}\ \emph {et~al.}(2015)\citenamefont {Reithmaier}, \citenamefont {Kaniber}, \citenamefont {Flassig}, \citenamefont {Lichtmannecker}, \citenamefont {M{\"u}ller}, \citenamefont {Andrejew}, \citenamefont {Vu{\v c}kovi{\'c}}, \citenamefont {Gross},\ and\ \citenamefont {Finley}}]{Reithmaier2015}%
  \BibitemOpen
  \bibfield  {author} {\bibinfo {author} {\bibfnamefont {G.}~\bibnamefont {Reithmaier}}, \bibinfo {author} {\bibfnamefont {M.}~\bibnamefont {Kaniber}}, \bibinfo {author} {\bibfnamefont {F.}~\bibnamefont {Flassig}}, \bibinfo {author} {\bibfnamefont {S.}~\bibnamefont {Lichtmannecker}}, \bibinfo {author} {\bibfnamefont {K.}~\bibnamefont {M{\"u}ller}}, \bibinfo {author} {\bibfnamefont {A.}~\bibnamefont {Andrejew}}, \bibinfo {author} {\bibfnamefont {J.}~\bibnamefont {Vu{\v c}kovi{\'c}}}, \bibinfo {author} {\bibfnamefont {R.}~\bibnamefont {Gross}},\ and\ \bibinfo {author} {\bibfnamefont {J.~J.}\ \bibnamefont {Finley}},\ }\bibfield  {title} {\bibinfo {title} {On-{{Chip Generation}}, {{Routing}}, and {{Detection}} of {{Resonance Fluorescence}}},\ }\href {https://doi.org/10.1021/acs.nanolett.5b01444} {\bibfield  {journal} {\bibinfo  {journal} {Nano Letters}\ }\textbf {\bibinfo {volume} {15}},\ \bibinfo {pages} {5208} (\bibinfo {year} {2015})}\BibitemShut {NoStop}%
\bibitem [{\citenamefont {Majety}\ \emph {et~al.}(2023)\citenamefont {Majety}, \citenamefont {Strohauer}, \citenamefont {Saha}, \citenamefont {Wietschorke}, \citenamefont {Finley}, \citenamefont {M{\"u}ller},\ and\ \citenamefont {Radulaski}}]{Majety2023}%
  \BibitemOpen
  \bibfield  {author} {\bibinfo {author} {\bibfnamefont {S.}~\bibnamefont {Majety}}, \bibinfo {author} {\bibfnamefont {S.}~\bibnamefont {Strohauer}}, \bibinfo {author} {\bibfnamefont {P.}~\bibnamefont {Saha}}, \bibinfo {author} {\bibfnamefont {F.}~\bibnamefont {Wietschorke}}, \bibinfo {author} {\bibfnamefont {J.~J.}\ \bibnamefont {Finley}}, \bibinfo {author} {\bibfnamefont {K.}~\bibnamefont {M{\"u}ller}},\ and\ \bibinfo {author} {\bibfnamefont {M.}~\bibnamefont {Radulaski}},\ }\bibfield  {title} {\bibinfo {title} {Triangular quantum photonic devices with integrated detectors in silicon carbide},\ }\href {https://doi.org/10.1088/2633-4356/acc302} {\bibfield  {journal} {\bibinfo  {journal} {Materials for Quantum Technology}\ }\textbf {\bibinfo {volume} {3}},\ \bibinfo {pages} {015004} (\bibinfo {year} {2023})}\BibitemShut {NoStop}%
\bibitem [{\citenamefont {Polakovic}\ \emph {et~al.}(2020)\citenamefont {Polakovic}, \citenamefont {Armstrong}, \citenamefont {Karapetrov}, \citenamefont {Meziani},\ and\ \citenamefont {Novosad}}]{Polakovic2020}%
  \BibitemOpen
  \bibfield  {author} {\bibinfo {author} {\bibfnamefont {T.}~\bibnamefont {Polakovic}}, \bibinfo {author} {\bibfnamefont {W.}~\bibnamefont {Armstrong}}, \bibinfo {author} {\bibfnamefont {G.}~\bibnamefont {Karapetrov}}, \bibinfo {author} {\bibfnamefont {Z.~E.}\ \bibnamefont {Meziani}},\ and\ \bibinfo {author} {\bibfnamefont {V.}~\bibnamefont {Novosad}},\ }\bibfield  {title} {\bibinfo {title} {Unconventional applications of superconducting nanowire single photon detectors},\ }\href {https://doi.org/10.3390/nano10061198} {\bibfield  {journal} {\bibinfo  {journal} {Nanomaterials}\ }\textbf {\bibinfo {volume} {10}},\ \bibinfo {pages} {1198} (\bibinfo {year} {2020})}\BibitemShut {NoStop}%
\bibitem [{\citenamefont {Shigefuji}\ \emph {et~al.}(2023)\citenamefont {Shigefuji}, \citenamefont {Osada}, \citenamefont {Yabuno}, \citenamefont {Miki}, \citenamefont {Terai},\ and\ \citenamefont {Noguchi}}]{Shigefuji2023}%
  \BibitemOpen
  \bibfield  {author} {\bibinfo {author} {\bibfnamefont {M.}~\bibnamefont {Shigefuji}}, \bibinfo {author} {\bibfnamefont {A.}~\bibnamefont {Osada}}, \bibinfo {author} {\bibfnamefont {M.}~\bibnamefont {Yabuno}}, \bibinfo {author} {\bibfnamefont {S.}~\bibnamefont {Miki}}, \bibinfo {author} {\bibfnamefont {H.}~\bibnamefont {Terai}},\ and\ \bibinfo {author} {\bibfnamefont {A.}~\bibnamefont {Noguchi}},\ }\href@noop {} {\bibinfo {title} {Efficient low-energy single-electron detection using a large-area superconducting microstrip}} (\bibinfo {year} {2023}),\ \Eprint {https://arxiv.org/abs/2301.11212} {arXiv:2301.11212 [cond-mat, physics:physics, physics:quant-ph]} \BibitemShut {NoStop}%
\bibitem [{\citenamefont {Hochberg}\ \emph {et~al.}(2019)\citenamefont {Hochberg}, \citenamefont {Charaev}, \citenamefont {Nam}, \citenamefont {Verma}, \citenamefont {Colangelo},\ and\ \citenamefont {Berggren}}]{Hochberg2019}%
  \BibitemOpen
  \bibfield  {author} {\bibinfo {author} {\bibfnamefont {Y.}~\bibnamefont {Hochberg}}, \bibinfo {author} {\bibfnamefont {I.}~\bibnamefont {Charaev}}, \bibinfo {author} {\bibfnamefont {S.-W.}\ \bibnamefont {Nam}}, \bibinfo {author} {\bibfnamefont {V.}~\bibnamefont {Verma}}, \bibinfo {author} {\bibfnamefont {M.}~\bibnamefont {Colangelo}},\ and\ \bibinfo {author} {\bibfnamefont {K.~K.}\ \bibnamefont {Berggren}},\ }\bibfield  {title} {\bibinfo {title} {Detecting {{Sub-GeV Dark Matter}} with {{Superconducting Nanowires}}},\ }\href {https://doi.org/10.1103/PhysRevLett.123.151802} {\bibfield  {journal} {\bibinfo  {journal} {Physical Review Letters}\ }\textbf {\bibinfo {volume} {123}},\ \bibinfo {pages} {151802} (\bibinfo {year} {2019})}\BibitemShut {NoStop}%
\bibitem [{\citenamefont {Chiles}\ \emph {et~al.}(2022)\citenamefont {Chiles}, \citenamefont {Charaev}, \citenamefont {Lasenby}, \citenamefont {Baryakhtar}, \citenamefont {Huang}, \citenamefont {Roshko}, \citenamefont {Burton}, \citenamefont {Colangelo}, \citenamefont {Van~Tilburg}, \citenamefont {Arvanitaki}, \citenamefont {Nam},\ and\ \citenamefont {Berggren}}]{Chiles2022}%
  \BibitemOpen
  \bibfield  {author} {\bibinfo {author} {\bibfnamefont {J.}~\bibnamefont {Chiles}}, \bibinfo {author} {\bibfnamefont {I.}~\bibnamefont {Charaev}}, \bibinfo {author} {\bibfnamefont {R.}~\bibnamefont {Lasenby}}, \bibinfo {author} {\bibfnamefont {M.}~\bibnamefont {Baryakhtar}}, \bibinfo {author} {\bibfnamefont {J.}~\bibnamefont {Huang}}, \bibinfo {author} {\bibfnamefont {A.}~\bibnamefont {Roshko}}, \bibinfo {author} {\bibfnamefont {G.}~\bibnamefont {Burton}}, \bibinfo {author} {\bibfnamefont {M.}~\bibnamefont {Colangelo}}, \bibinfo {author} {\bibfnamefont {K.}~\bibnamefont {Van~Tilburg}}, \bibinfo {author} {\bibfnamefont {A.}~\bibnamefont {Arvanitaki}}, \bibinfo {author} {\bibfnamefont {S.~W.}\ \bibnamefont {Nam}},\ and\ \bibinfo {author} {\bibfnamefont {K.~K.}\ \bibnamefont {Berggren}},\ }\bibfield  {title} {\bibinfo {title} {New {{Constraints}} on {{Dark Photon Dark Matter}} with {{Superconducting Nanowire Detectors}} in an {{Optical Haloscope}}},\ }\href {https://doi.org/10.1103/PhysRevLett.128.231802}
  {\bibfield  {journal} {\bibinfo  {journal} {Physical Review Letters}\ }\textbf {\bibinfo {volume} {128}},\ \bibinfo {pages} {231802} (\bibinfo {year} {2022})}\BibitemShut {NoStop}%
\bibitem [{\citenamefont {Clem}\ and\ \citenamefont {Berggren}(2011)}]{Clem2011}%
  \BibitemOpen
  \bibfield  {author} {\bibinfo {author} {\bibfnamefont {J.~R.}\ \bibnamefont {Clem}}\ and\ \bibinfo {author} {\bibfnamefont {K.~K.}\ \bibnamefont {Berggren}},\ }\bibfield  {title} {\bibinfo {title} {Geometry-dependent critical currents in superconducting nanocircuits},\ }\bibfield  {journal} {\bibinfo  {journal} {Physical Review B}\ }\textbf {\bibinfo {volume} {84}},\ \href {https://doi.org/10.1103/PhysRevB.84.174510} {10.1103/PhysRevB.84.174510} (\bibinfo {year} {2011})\BibitemShut {NoStop}%
\bibitem [{\citenamefont {J{\"o}nsson}\ \emph {et~al.}(2022)\citenamefont {J{\"o}nsson}, \citenamefont {Vedin}, \citenamefont {Gyger}, \citenamefont {Sutton}, \citenamefont {Steinhauer}, \citenamefont {Zwiller}, \citenamefont {Wallin},\ and\ \citenamefont {Lidmar}}]{Jonsson2022}%
  \BibitemOpen
  \bibfield  {author} {\bibinfo {author} {\bibfnamefont {M.}~\bibnamefont {J{\"o}nsson}}, \bibinfo {author} {\bibfnamefont {R.}~\bibnamefont {Vedin}}, \bibinfo {author} {\bibfnamefont {S.}~\bibnamefont {Gyger}}, \bibinfo {author} {\bibfnamefont {J.~A.}\ \bibnamefont {Sutton}}, \bibinfo {author} {\bibfnamefont {S.}~\bibnamefont {Steinhauer}}, \bibinfo {author} {\bibfnamefont {V.}~\bibnamefont {Zwiller}}, \bibinfo {author} {\bibfnamefont {M.}~\bibnamefont {Wallin}},\ and\ \bibinfo {author} {\bibfnamefont {J.}~\bibnamefont {Lidmar}},\ }\bibfield  {title} {\bibinfo {title} {Current {{Crowding}} in {{Nanoscale Superconductors}} within the {{Ginzburg-Landau Model}}},\ }\href {https://doi.org/10.1103/PhysRevApplied.17.064046} {\bibfield  {journal} {\bibinfo  {journal} {Physical Review Applied}\ }\textbf {\bibinfo {volume} {17}},\ \bibinfo {pages} {064046} (\bibinfo {year} {2022})}\BibitemShut {NoStop}%
\bibitem [{\citenamefont {Akhlaghi}\ \emph {et~al.}(2012)\citenamefont {Akhlaghi}, \citenamefont {Atikian}, \citenamefont {Eftekharian}, \citenamefont {Loncar},\ and\ \citenamefont {Majedi}}]{Akhlaghi2012}%
  \BibitemOpen
  \bibfield  {author} {\bibinfo {author} {\bibfnamefont {M.~K.}\ \bibnamefont {Akhlaghi}}, \bibinfo {author} {\bibfnamefont {H.}~\bibnamefont {Atikian}}, \bibinfo {author} {\bibfnamefont {A.}~\bibnamefont {Eftekharian}}, \bibinfo {author} {\bibfnamefont {M.}~\bibnamefont {Loncar}},\ and\ \bibinfo {author} {\bibfnamefont {A.~H.}\ \bibnamefont {Majedi}},\ }\bibfield  {title} {\bibinfo {title} {Reduced dark counts in optimized geometries for superconducting nanowire single photon detectors},\ }\href {https://doi.org/10.1364/oe.20.023610} {\bibfield  {journal} {\bibinfo  {journal} {Optics Express}\ }\textbf {\bibinfo {volume} {20}},\ \bibinfo {pages} {23610} (\bibinfo {year} {2012})}\BibitemShut {NoStop}%
\bibitem [{\citenamefont {Henrich}\ \emph {et~al.}(2013)\citenamefont {Henrich}, \citenamefont {Rehm}, \citenamefont {Dorner}, \citenamefont {Hofherr}, \citenamefont {Il'in}, \citenamefont {Semenov},\ and\ \citenamefont {Siegel}}]{Henrich2013}%
  \BibitemOpen
  \bibfield  {author} {\bibinfo {author} {\bibfnamefont {D.}~\bibnamefont {Henrich}}, \bibinfo {author} {\bibfnamefont {L.}~\bibnamefont {Rehm}}, \bibinfo {author} {\bibfnamefont {S.}~\bibnamefont {Dorner}}, \bibinfo {author} {\bibfnamefont {M.}~\bibnamefont {Hofherr}}, \bibinfo {author} {\bibfnamefont {K.}~\bibnamefont {Il'in}}, \bibinfo {author} {\bibfnamefont {A.}~\bibnamefont {Semenov}},\ and\ \bibinfo {author} {\bibfnamefont {M.}~\bibnamefont {Siegel}},\ }\bibfield  {title} {\bibinfo {title} {Detection {{Efficiency}} of a {{Spiral-Nanowire Superconducting Single-Photon Detector}}},\ }\href {https://doi.org/10.1109/TASC.2013.2237936} {\bibfield  {journal} {\bibinfo  {journal} {IEEE Transactions on Applied Superconductivity}\ }\textbf {\bibinfo {volume} {23}},\ \bibinfo {pages} {2200405} (\bibinfo {year} {2013})}\BibitemShut {NoStop}%
\bibitem [{\citenamefont {Charaev}\ \emph {et~al.}(2017)\citenamefont {Charaev}, \citenamefont {Semenov}, \citenamefont {Doerner}, \citenamefont {Gomard}, \citenamefont {Ilin},\ and\ \citenamefont {Siegel}}]{Charaev2017}%
  \BibitemOpen
  \bibfield  {author} {\bibinfo {author} {\bibfnamefont {I.}~\bibnamefont {Charaev}}, \bibinfo {author} {\bibfnamefont {A.}~\bibnamefont {Semenov}}, \bibinfo {author} {\bibfnamefont {S.}~\bibnamefont {Doerner}}, \bibinfo {author} {\bibfnamefont {G.}~\bibnamefont {Gomard}}, \bibinfo {author} {\bibfnamefont {K.}~\bibnamefont {Ilin}},\ and\ \bibinfo {author} {\bibfnamefont {M.}~\bibnamefont {Siegel}},\ }\bibfield  {title} {\bibinfo {title} {Current dependence of the hot-spot response spectrum of superconducting single-photon detectors with different layouts},\ }\href {https://doi.org/10.1088/1361-6668/30/2/025016} {\bibfield  {journal} {\bibinfo  {journal} {Superconductor Science and Technology}\ }\textbf {\bibinfo {volume} {30}},\ \bibinfo {pages} {025016} (\bibinfo {year} {2017})}\BibitemShut {NoStop}%
\bibitem [{\citenamefont {Baghdadi}\ \emph {et~al.}(2021)\citenamefont {Baghdadi}, \citenamefont {Schmidt}, \citenamefont {Jahani}, \citenamefont {Charaev}, \citenamefont {M{\"u}ller}, \citenamefont {Colangelo}, \citenamefont {Zhu}, \citenamefont {Ilin}, \citenamefont {Semenov}, \citenamefont {Jacob}, \citenamefont {Siegel},\ and\ \citenamefont {Berggren}}]{Baghdadi2021}%
  \BibitemOpen
  \bibfield  {author} {\bibinfo {author} {\bibfnamefont {R.}~\bibnamefont {Baghdadi}}, \bibinfo {author} {\bibfnamefont {E.}~\bibnamefont {Schmidt}}, \bibinfo {author} {\bibfnamefont {S.}~\bibnamefont {Jahani}}, \bibinfo {author} {\bibfnamefont {I.}~\bibnamefont {Charaev}}, \bibinfo {author} {\bibfnamefont {M.~G.~W.}\ \bibnamefont {M{\"u}ller}}, \bibinfo {author} {\bibfnamefont {M.}~\bibnamefont {Colangelo}}, \bibinfo {author} {\bibfnamefont {D.}~\bibnamefont {Zhu}}, \bibinfo {author} {\bibfnamefont {K.}~\bibnamefont {Ilin}}, \bibinfo {author} {\bibfnamefont {A.~D.}\ \bibnamefont {Semenov}}, \bibinfo {author} {\bibfnamefont {Z.}~\bibnamefont {Jacob}}, \bibinfo {author} {\bibfnamefont {M.}~\bibnamefont {Siegel}},\ and\ \bibinfo {author} {\bibfnamefont {K.~K.}\ \bibnamefont {Berggren}},\ }\bibfield  {title} {\bibinfo {title} {Enhancing the performance of superconducting nanowire-based detectors with high-filling factor by using variable thickness},\ }\href {https://doi.org/10.1088/1361-6668/abdba6} {\bibfield
  {journal} {\bibinfo  {journal} {Superconductor Science and Technology}\ }\textbf {\bibinfo {volume} {34}},\ \bibinfo {pages} {035010} (\bibinfo {year} {2021})}\BibitemShut {NoStop}%
\bibitem [{\citenamefont {Xiong}\ \emph {et~al.}(2022)\citenamefont {Xiong}, \citenamefont {Zhang}, \citenamefont {Xu}, \citenamefont {You}, \citenamefont {Zhang}, \citenamefont {Zhang}, \citenamefont {Zhang}, \citenamefont {Fan}, \citenamefont {Wang}, \citenamefont {Li},\ and\ \citenamefont {Wang}}]{Xiong2022}%
  \BibitemOpen
  \bibfield  {author} {\bibinfo {author} {\bibfnamefont {J.-M.}\ \bibnamefont {Xiong}}, \bibinfo {author} {\bibfnamefont {W.-J.}\ \bibnamefont {Zhang}}, \bibinfo {author} {\bibfnamefont {G.-Z.}\ \bibnamefont {Xu}}, \bibinfo {author} {\bibfnamefont {L.-X.}\ \bibnamefont {You}}, \bibinfo {author} {\bibfnamefont {X.-Y.}\ \bibnamefont {Zhang}}, \bibinfo {author} {\bibfnamefont {L.}~\bibnamefont {Zhang}}, \bibinfo {author} {\bibfnamefont {C.-J.}\ \bibnamefont {Zhang}}, \bibinfo {author} {\bibfnamefont {D.-H.}\ \bibnamefont {Fan}}, \bibinfo {author} {\bibfnamefont {Y.-Z.}\ \bibnamefont {Wang}}, \bibinfo {author} {\bibfnamefont {H.}~\bibnamefont {Li}},\ and\ \bibinfo {author} {\bibfnamefont {Z.}~\bibnamefont {Wang}},\ }\bibfield  {title} {\bibinfo {title} {Reducing current crowding in meander superconducting strip single-photon detectors by thickening bends},\ }\href {https://doi.org/10.1088/1361-6668/ac5fe4} {\bibfield  {journal} {\bibinfo  {journal} {Superconductor Science and Technology}\ }\textbf {\bibinfo
  {volume} {35}},\ \bibinfo {pages} {055015} (\bibinfo {year} {2022})}\BibitemShut {NoStop}%
\bibitem [{\citenamefont {Henrich}\ \emph {et~al.}(2012)\citenamefont {Henrich}, \citenamefont {Reichensperger}, \citenamefont {Hofherr}, \citenamefont {Meckbach}, \citenamefont {Il'in}, \citenamefont {Siegel}, \citenamefont {Semenov}, \citenamefont {Zotova},\ and\ \citenamefont {Vodolazov}}]{Henrich2012}%
  \BibitemOpen
  \bibfield  {author} {\bibinfo {author} {\bibfnamefont {D.}~\bibnamefont {Henrich}}, \bibinfo {author} {\bibfnamefont {P.}~\bibnamefont {Reichensperger}}, \bibinfo {author} {\bibfnamefont {M.}~\bibnamefont {Hofherr}}, \bibinfo {author} {\bibfnamefont {J.~M.}\ \bibnamefont {Meckbach}}, \bibinfo {author} {\bibfnamefont {K.}~\bibnamefont {Il'in}}, \bibinfo {author} {\bibfnamefont {M.}~\bibnamefont {Siegel}}, \bibinfo {author} {\bibfnamefont {A.}~\bibnamefont {Semenov}}, \bibinfo {author} {\bibfnamefont {A.}~\bibnamefont {Zotova}},\ and\ \bibinfo {author} {\bibfnamefont {D.~{\relax Yu}.}\ \bibnamefont {Vodolazov}},\ }\bibfield  {title} {\bibinfo {title} {Geometry-induced reduction of the critical current in superconducting nanowires},\ }\href {https://doi.org/10.1103/PhysRevB.86.144504} {\bibfield  {journal} {\bibinfo  {journal} {Physical Review B}\ }\textbf {\bibinfo {volume} {86}},\ \bibinfo {pages} {144504} (\bibinfo {year} {2012})}\BibitemShut {NoStop}%
\bibitem [{\citenamefont {Hortensius}\ \emph {et~al.}(2012)\citenamefont {Hortensius}, \citenamefont {Driessen}, \citenamefont {Klapwijk}, \citenamefont {Berggren},\ and\ \citenamefont {Clem}}]{Hortensius2012}%
  \BibitemOpen
  \bibfield  {author} {\bibinfo {author} {\bibfnamefont {H.~L.}\ \bibnamefont {Hortensius}}, \bibinfo {author} {\bibfnamefont {E.~F.~C.}\ \bibnamefont {Driessen}}, \bibinfo {author} {\bibfnamefont {T.~M.}\ \bibnamefont {Klapwijk}}, \bibinfo {author} {\bibfnamefont {K.~K.}\ \bibnamefont {Berggren}},\ and\ \bibinfo {author} {\bibfnamefont {J.~R.}\ \bibnamefont {Clem}},\ }\bibfield  {title} {\bibinfo {title} {Critical-current reduction in thin superconducting wires due to current crowding},\ }\href {https://doi.org/10.1063/1.4711217} {\bibfield  {journal} {\bibinfo  {journal} {Applied Physics Letters}\ }\textbf {\bibinfo {volume} {100}},\ \bibinfo {pages} {182602} (\bibinfo {year} {2012})}\BibitemShut {NoStop}%
\bibitem [{\citenamefont {Frasca}\ \emph {et~al.}(2019)\citenamefont {Frasca}, \citenamefont {Korzh}, \citenamefont {Colangelo}, \citenamefont {Zhu}, \citenamefont {Lita}, \citenamefont {Allmaras}, \citenamefont {Wollman}, \citenamefont {Verma}, \citenamefont {Dane}, \citenamefont {Ramirez}, \citenamefont {Beyer}, \citenamefont {Nam}, \citenamefont {Kozorezov}, \citenamefont {Shaw},\ and\ \citenamefont {Berggren}}]{Frasca2019}%
  \BibitemOpen
  \bibfield  {author} {\bibinfo {author} {\bibfnamefont {S.}~\bibnamefont {Frasca}}, \bibinfo {author} {\bibfnamefont {B.}~\bibnamefont {Korzh}}, \bibinfo {author} {\bibfnamefont {M.}~\bibnamefont {Colangelo}}, \bibinfo {author} {\bibfnamefont {D.}~\bibnamefont {Zhu}}, \bibinfo {author} {\bibfnamefont {A.~E.}\ \bibnamefont {Lita}}, \bibinfo {author} {\bibfnamefont {J.~P.}\ \bibnamefont {Allmaras}}, \bibinfo {author} {\bibfnamefont {E.~E.}\ \bibnamefont {Wollman}}, \bibinfo {author} {\bibfnamefont {V.~B.}\ \bibnamefont {Verma}}, \bibinfo {author} {\bibfnamefont {A.~E.}\ \bibnamefont {Dane}}, \bibinfo {author} {\bibfnamefont {E.}~\bibnamefont {Ramirez}}, \bibinfo {author} {\bibfnamefont {A.~D.}\ \bibnamefont {Beyer}}, \bibinfo {author} {\bibfnamefont {S.~W.}\ \bibnamefont {Nam}}, \bibinfo {author} {\bibfnamefont {A.~G.}\ \bibnamefont {Kozorezov}}, \bibinfo {author} {\bibfnamefont {M.~D.}\ \bibnamefont {Shaw}},\ and\ \bibinfo {author} {\bibfnamefont {K.~K.}\ \bibnamefont {Berggren}},\ }\bibfield  {title}
  {\bibinfo {title} {Determining the depairing current in superconducting nanowire single-photon detectors},\ }\href {https://doi.org/10.1103/PhysRevB.100.054520} {\bibfield  {journal} {\bibinfo  {journal} {Physical Review B}\ }\textbf {\bibinfo {volume} {100}},\ \bibinfo {pages} {054520} (\bibinfo {year} {2019})}\BibitemShut {NoStop}%
\bibitem [{\citenamefont {Semenov}\ \emph {et~al.}(2015)\citenamefont {Semenov}, \citenamefont {Charaev}, \citenamefont {Lusche}, \citenamefont {Ilin}, \citenamefont {Siegel}, \citenamefont {H{\"u}bers}, \citenamefont {Bralovi{\'c}}, \citenamefont {Dopf},\ and\ \citenamefont {Vodolazov}}]{Semenov2015}%
  \BibitemOpen
  \bibfield  {author} {\bibinfo {author} {\bibfnamefont {A.}~\bibnamefont {Semenov}}, \bibinfo {author} {\bibfnamefont {I.}~\bibnamefont {Charaev}}, \bibinfo {author} {\bibfnamefont {R.}~\bibnamefont {Lusche}}, \bibinfo {author} {\bibfnamefont {K.}~\bibnamefont {Ilin}}, \bibinfo {author} {\bibfnamefont {M.}~\bibnamefont {Siegel}}, \bibinfo {author} {\bibfnamefont {H.-W.}\ \bibnamefont {H{\"u}bers}}, \bibinfo {author} {\bibfnamefont {N.}~\bibnamefont {Bralovi{\'c}}}, \bibinfo {author} {\bibfnamefont {K.}~\bibnamefont {Dopf}},\ and\ \bibinfo {author} {\bibfnamefont {D.~Y.}\ \bibnamefont {Vodolazov}},\ }\bibfield  {title} {\bibinfo {title} {Asymmetry in the effect of magnetic field on photon detection and dark counts in bended nanostrips},\ }\href {https://doi.org/10.1103/PhysRevB.92.174518} {\bibfield  {journal} {\bibinfo  {journal} {Physical Review B}\ }\textbf {\bibinfo {volume} {92}},\ \bibinfo {pages} {174518} (\bibinfo {year} {2015})}\BibitemShut {NoStop}%
\bibitem [{\citenamefont {Zhang}\ \emph {et~al.}(2014)\citenamefont {Zhang}, \citenamefont {You}, \citenamefont {Liu}, \citenamefont {Zhang}, \citenamefont {Zhang}, \citenamefont {Liu}, \citenamefont {Wu}, \citenamefont {He}, \citenamefont {Lv}, \citenamefont {Wang},\ and\ \citenamefont {Xie}}]{Zhang2014}%
  \BibitemOpen
  \bibfield  {author} {\bibinfo {author} {\bibfnamefont {L.}~\bibnamefont {Zhang}}, \bibinfo {author} {\bibfnamefont {L.}~\bibnamefont {You}}, \bibinfo {author} {\bibfnamefont {D.}~\bibnamefont {Liu}}, \bibinfo {author} {\bibfnamefont {W.}~\bibnamefont {Zhang}}, \bibinfo {author} {\bibfnamefont {L.}~\bibnamefont {Zhang}}, \bibinfo {author} {\bibfnamefont {X.}~\bibnamefont {Liu}}, \bibinfo {author} {\bibfnamefont {J.}~\bibnamefont {Wu}}, \bibinfo {author} {\bibfnamefont {Y.}~\bibnamefont {He}}, \bibinfo {author} {\bibfnamefont {C.}~\bibnamefont {Lv}}, \bibinfo {author} {\bibfnamefont {Z.}~\bibnamefont {Wang}},\ and\ \bibinfo {author} {\bibfnamefont {X.}~\bibnamefont {Xie}},\ }\bibfield  {title} {\bibinfo {title} {Characterization of superconducting nanowire single-photon detector with artificial constrictions},\ }\href {https://doi.org/10.1063/1.4881981} {\bibfield  {journal} {\bibinfo  {journal} {AIP Advances}\ }\textbf {\bibinfo {volume} {4}},\ \bibinfo {pages} {067114} (\bibinfo {year} {2014})}\BibitemShut
  {NoStop}%
\bibitem [{\citenamefont {Zhang}\ \emph {et~al.}(2022)\citenamefont {Zhang}, \citenamefont {Zhang}, \citenamefont {Huang}, \citenamefont {Yang}, \citenamefont {You}, \citenamefont {Liu}, \citenamefont {Hu}, \citenamefont {Xiao}, \citenamefont {Zhang}, \citenamefont {Wang}, \citenamefont {Li}, \citenamefont {Wang},\ and\ \citenamefont {Li}}]{Zhang2022b}%
  \BibitemOpen
  \bibfield  {author} {\bibinfo {author} {\bibfnamefont {X.}~\bibnamefont {Zhang}}, \bibinfo {author} {\bibfnamefont {X.}~\bibnamefont {Zhang}}, \bibinfo {author} {\bibfnamefont {J.}~\bibnamefont {Huang}}, \bibinfo {author} {\bibfnamefont {C.}~\bibnamefont {Yang}}, \bibinfo {author} {\bibfnamefont {L.}~\bibnamefont {You}}, \bibinfo {author} {\bibfnamefont {X.}~\bibnamefont {Liu}}, \bibinfo {author} {\bibfnamefont {P.}~\bibnamefont {Hu}}, \bibinfo {author} {\bibfnamefont {Y.}~\bibnamefont {Xiao}}, \bibinfo {author} {\bibfnamefont {W.}~\bibnamefont {Zhang}}, \bibinfo {author} {\bibfnamefont {Y.}~\bibnamefont {Wang}}, \bibinfo {author} {\bibfnamefont {L.}~\bibnamefont {Li}}, \bibinfo {author} {\bibfnamefont {Z.}~\bibnamefont {Wang}},\ and\ \bibinfo {author} {\bibfnamefont {H.}~\bibnamefont {Li}},\ }\bibfield  {title} {\bibinfo {title} {Geometric origin of intrinsic dark counts in superconducting nanowire single-photon detectors},\ }\href {https://doi.org/10.1016/j.supcon.2022.100006} {\bibfield  {journal}
  {\bibinfo  {journal} {Superconductivity}\ }\textbf {\bibinfo {volume} {1}},\ \bibinfo {pages} {100006} (\bibinfo {year} {2022})}\BibitemShut {NoStop}%
\bibitem [{\citenamefont {Zhang}\ \emph {et~al.}(2019)\citenamefont {Zhang}, \citenamefont {Jia}, \citenamefont {You}, \citenamefont {Ou}, \citenamefont {Huang}, \citenamefont {Zhang}, \citenamefont {Li}, \citenamefont {Wang},\ and\ \citenamefont {Xie}}]{Zhang2019}%
  \BibitemOpen
  \bibfield  {author} {\bibinfo {author} {\bibfnamefont {W.}~\bibnamefont {Zhang}}, \bibinfo {author} {\bibfnamefont {Q.}~\bibnamefont {Jia}}, \bibinfo {author} {\bibfnamefont {L.}~\bibnamefont {You}}, \bibinfo {author} {\bibfnamefont {X.}~\bibnamefont {Ou}}, \bibinfo {author} {\bibfnamefont {H.}~\bibnamefont {Huang}}, \bibinfo {author} {\bibfnamefont {L.}~\bibnamefont {Zhang}}, \bibinfo {author} {\bibfnamefont {H.}~\bibnamefont {Li}}, \bibinfo {author} {\bibfnamefont {Z.}~\bibnamefont {Wang}},\ and\ \bibinfo {author} {\bibfnamefont {X.}~\bibnamefont {Xie}},\ }\bibfield  {title} {\bibinfo {title} {Saturating {{Intrinsic Detection Efficiency}} of {{Superconducting Nanowire Single-Photon Detectors}} via {{Defect Engineering}}},\ }\href {https://doi.org/10.1103/PhysRevApplied.12.044040} {\bibfield  {journal} {\bibinfo  {journal} {Physical Review Applied}\ }\textbf {\bibinfo {volume} {12}},\ \bibinfo {pages} {044040} (\bibinfo {year} {2019})}\BibitemShut {NoStop}%
\bibitem [{\citenamefont {Strohauer}\ \emph {et~al.}(2023)\citenamefont {Strohauer}, \citenamefont {Wietschorke}, \citenamefont {Zugliani}, \citenamefont {Flaschmann}, \citenamefont {Schmid}, \citenamefont {Grotowski}, \citenamefont {M{\"u}ller}, \citenamefont {Jonas}, \citenamefont {Althammer}, \citenamefont {Gross}, \citenamefont {M{\"u}ller},\ and\ \citenamefont {Finley}}]{Strohauer2023}%
  \BibitemOpen
  \bibfield  {author} {\bibinfo {author} {\bibfnamefont {S.}~\bibnamefont {Strohauer}}, \bibinfo {author} {\bibfnamefont {F.}~\bibnamefont {Wietschorke}}, \bibinfo {author} {\bibfnamefont {L.}~\bibnamefont {Zugliani}}, \bibinfo {author} {\bibfnamefont {R.}~\bibnamefont {Flaschmann}}, \bibinfo {author} {\bibfnamefont {C.}~\bibnamefont {Schmid}}, \bibinfo {author} {\bibfnamefont {S.}~\bibnamefont {Grotowski}}, \bibinfo {author} {\bibfnamefont {M.}~\bibnamefont {M{\"u}ller}}, \bibinfo {author} {\bibfnamefont {B.}~\bibnamefont {Jonas}}, \bibinfo {author} {\bibfnamefont {M.}~\bibnamefont {Althammer}}, \bibinfo {author} {\bibfnamefont {R.}~\bibnamefont {Gross}}, \bibinfo {author} {\bibfnamefont {K.}~\bibnamefont {M{\"u}ller}},\ and\ \bibinfo {author} {\bibfnamefont {J.~J.}\ \bibnamefont {Finley}},\ }\bibfield  {title} {\bibinfo {title} {Site-{{Selective Enhancement}} of {{Superconducting Nanowire Single}}-{{Photon Detectors}} via {{Local Helium Ion Irradiation}}},\ }\href {https://doi.org/10.1002/qute.202300139}
  {\bibfield  {journal} {\bibinfo  {journal} {Advanced Quantum Technologies}\ }\textbf {\bibinfo {volume} {6}},\ \bibinfo {pages} {2300139} (\bibinfo {year} {2023})}\BibitemShut {NoStop}%
\bibitem [{\citenamefont {Brenner}\ \emph {et~al.}(2012)\citenamefont {Brenner}, \citenamefont {Roy}, \citenamefont {Shah},\ and\ \citenamefont {Bezryadin}}]{Brenner2012}%
  \BibitemOpen
  \bibfield  {author} {\bibinfo {author} {\bibfnamefont {M.~W.}\ \bibnamefont {Brenner}}, \bibinfo {author} {\bibfnamefont {D.}~\bibnamefont {Roy}}, \bibinfo {author} {\bibfnamefont {N.}~\bibnamefont {Shah}},\ and\ \bibinfo {author} {\bibfnamefont {A.}~\bibnamefont {Bezryadin}},\ }\bibfield  {title} {\bibinfo {title} {Dynamics of superconducting nanowires shunted with an external resistor},\ }\href {https://doi.org/10.1103/PhysRevB.85.224507} {\bibfield  {journal} {\bibinfo  {journal} {Physical Review B}\ }\textbf {\bibinfo {volume} {85}},\ \bibinfo {pages} {224507} (\bibinfo {year} {2012})}\BibitemShut {NoStop}%
\bibitem [{\citenamefont {Charaev}\ \emph {et~al.}(2024)\citenamefont {Charaev}, \citenamefont {Batson}, \citenamefont {Cherednichenko}, \citenamefont {Reidy}, \citenamefont {Drakinskiy}, \citenamefont {Yu}, \citenamefont {{Lara-Avila}}, \citenamefont {Thomsen}, \citenamefont {Colangelo}, \citenamefont {Incalza}, \citenamefont {Ilin}, \citenamefont {Schilling},\ and\ \citenamefont {Berggren}}]{Charaev2024}%
  \BibitemOpen
  \bibfield  {author} {\bibinfo {author} {\bibfnamefont {I.}~\bibnamefont {Charaev}}, \bibinfo {author} {\bibfnamefont {E.~K.}\ \bibnamefont {Batson}}, \bibinfo {author} {\bibfnamefont {S.}~\bibnamefont {Cherednichenko}}, \bibinfo {author} {\bibfnamefont {K.}~\bibnamefont {Reidy}}, \bibinfo {author} {\bibfnamefont {V.}~\bibnamefont {Drakinskiy}}, \bibinfo {author} {\bibfnamefont {Y.}~\bibnamefont {Yu}}, \bibinfo {author} {\bibfnamefont {S.}~\bibnamefont {{Lara-Avila}}}, \bibinfo {author} {\bibfnamefont {J.~D.}\ \bibnamefont {Thomsen}}, \bibinfo {author} {\bibfnamefont {M.}~\bibnamefont {Colangelo}}, \bibinfo {author} {\bibfnamefont {F.}~\bibnamefont {Incalza}}, \bibinfo {author} {\bibfnamefont {K.}~\bibnamefont {Ilin}}, \bibinfo {author} {\bibfnamefont {A.}~\bibnamefont {Schilling}},\ and\ \bibinfo {author} {\bibfnamefont {K.~K.}\ \bibnamefont {Berggren}},\ }\bibfield  {title} {\bibinfo {title} {Single-photon detection using large-scale high-temperature {{MgB2}} sensors at 20 {{K}}},\ }\href
  {https://doi.org/10.1038/s41467-024-47353-x} {\bibfield  {journal} {\bibinfo  {journal} {Nature Communications}\ }\textbf {\bibinfo {volume} {15}},\ \bibinfo {pages} {3973} (\bibinfo {year} {2024})}\BibitemShut {NoStop}%
\bibitem [{\citenamefont {Wang}\ \emph {et~al.}(2024)\citenamefont {Wang}, \citenamefont {Zhang}, \citenamefont {Zhang}, \citenamefont {Xu}, \citenamefont {Xiong}, \citenamefont {Chen}, \citenamefont {Hong}, \citenamefont {Liu}, \citenamefont {Yuan}, \citenamefont {Wu}, \citenamefont {Wang},\ and\ \citenamefont {You}}]{Wang2024}%
  \BibitemOpen
  \bibfield  {author} {\bibinfo {author} {\bibfnamefont {Y.-Z.}\ \bibnamefont {Wang}}, \bibinfo {author} {\bibfnamefont {W.-J.}\ \bibnamefont {Zhang}}, \bibinfo {author} {\bibfnamefont {X.-Y.}\ \bibnamefont {Zhang}}, \bibinfo {author} {\bibfnamefont {G.-Z.}\ \bibnamefont {Xu}}, \bibinfo {author} {\bibfnamefont {J.-M.}\ \bibnamefont {Xiong}}, \bibinfo {author} {\bibfnamefont {Z.-G.}\ \bibnamefont {Chen}}, \bibinfo {author} {\bibfnamefont {Y.-Y.}\ \bibnamefont {Hong}}, \bibinfo {author} {\bibfnamefont {X.-Y.}\ \bibnamefont {Liu}}, \bibinfo {author} {\bibfnamefont {P.-S.}\ \bibnamefont {Yuan}}, \bibinfo {author} {\bibfnamefont {L.}~\bibnamefont {Wu}}, \bibinfo {author} {\bibfnamefont {Z.}~\bibnamefont {Wang}},\ and\ \bibinfo {author} {\bibfnamefont {L.-X.}\ \bibnamefont {You}},\ }\bibfield  {title} {\bibinfo {title} {Free-space coupled, large-active-area superconducting microstrip single-photon detector for photon-counting time-of-flight imaging},\ }\href {https://doi.org/10.1364/AO.519475} {\bibfield  {journal}
  {\bibinfo  {journal} {Applied Optics}\ }\textbf {\bibinfo {volume} {63}},\ \bibinfo {pages} {3130} (\bibinfo {year} {2024})}\BibitemShut {NoStop}%
\bibitem [{\citenamefont {Kahl}\ \emph {et~al.}(2015)\citenamefont {Kahl}, \citenamefont {Ferrari}, \citenamefont {Kovalyuk}, \citenamefont {Goltsman}, \citenamefont {Korneev},\ and\ \citenamefont {Pernice}}]{Kahl2015}%
  \BibitemOpen
  \bibfield  {author} {\bibinfo {author} {\bibfnamefont {O.}~\bibnamefont {Kahl}}, \bibinfo {author} {\bibfnamefont {S.}~\bibnamefont {Ferrari}}, \bibinfo {author} {\bibfnamefont {V.}~\bibnamefont {Kovalyuk}}, \bibinfo {author} {\bibfnamefont {G.~N.}\ \bibnamefont {Goltsman}}, \bibinfo {author} {\bibfnamefont {A.}~\bibnamefont {Korneev}},\ and\ \bibinfo {author} {\bibfnamefont {W.~H.~P.}\ \bibnamefont {Pernice}},\ }\bibfield  {title} {\bibinfo {title} {Waveguide integrated superconducting single-photon detectors with high internal quantum efficiency at telecom wavelengths},\ }\href {https://doi.org/10.1038/srep10941} {\bibfield  {journal} {\bibinfo  {journal} {Scientific Reports}\ }\textbf {\bibinfo {volume} {5}},\ \bibinfo {pages} {10941} (\bibinfo {year} {2015})}\BibitemShut {NoStop}%
\bibitem [{\citenamefont {Shibata}\ \emph {et~al.}(2013)\citenamefont {Shibata}, \citenamefont {Shimizu}, \citenamefont {Takesue},\ and\ \citenamefont {Tokura}}]{Shibata2013}%
  \BibitemOpen
  \bibfield  {author} {\bibinfo {author} {\bibfnamefont {H.}~\bibnamefont {Shibata}}, \bibinfo {author} {\bibfnamefont {K.}~\bibnamefont {Shimizu}}, \bibinfo {author} {\bibfnamefont {H.}~\bibnamefont {Takesue}},\ and\ \bibinfo {author} {\bibfnamefont {Y.}~\bibnamefont {Tokura}},\ }\bibfield  {title} {\bibinfo {title} {Superconducting {{Nanowire Single-Photon Detector}} with {{Ultralow Dark Count Rate Using Cold Optical Filters}}},\ }\href {https://doi.org/10.7567/APEX.6.072801} {\bibfield  {journal} {\bibinfo  {journal} {Applied Physics Express}\ }\textbf {\bibinfo {volume} {6}},\ \bibinfo {pages} {072801} (\bibinfo {year} {2013})}\BibitemShut {NoStop}%
\bibitem [{\citenamefont {Kerman}\ \emph {et~al.}(2013)\citenamefont {Kerman}, \citenamefont {Rosenberg}, \citenamefont {Molnar},\ and\ \citenamefont {Dauler}}]{Kerman2013}%
  \BibitemOpen
  \bibfield  {author} {\bibinfo {author} {\bibfnamefont {A.~J.}\ \bibnamefont {Kerman}}, \bibinfo {author} {\bibfnamefont {D.}~\bibnamefont {Rosenberg}}, \bibinfo {author} {\bibfnamefont {R.~J.}\ \bibnamefont {Molnar}},\ and\ \bibinfo {author} {\bibfnamefont {E.~A.}\ \bibnamefont {Dauler}},\ }\bibfield  {title} {\bibinfo {title} {Readout of superconducting nanowire single-photon detectors at high count rates},\ }\href {https://doi.org/10.1063/1.4799397} {\bibfield  {journal} {\bibinfo  {journal} {Journal of Applied Physics}\ }\textbf {\bibinfo {volume} {113}},\ \bibinfo {pages} {144511} (\bibinfo {year} {2013})}\BibitemShut {NoStop}%
\bibitem [{\citenamefont {Flaschmann}\ \emph {et~al.}(2023)\citenamefont {Flaschmann}, \citenamefont {Zugliani}, \citenamefont {Schmid}, \citenamefont {Spedicato}, \citenamefont {Strohauer}, \citenamefont {Wietschorke}, \citenamefont {Flassig}, \citenamefont {Finley},\ and\ \citenamefont {M{\"u}ller}}]{Flaschmann2023}%
  \BibitemOpen
  \bibfield  {author} {\bibinfo {author} {\bibfnamefont {R.}~\bibnamefont {Flaschmann}}, \bibinfo {author} {\bibfnamefont {L.}~\bibnamefont {Zugliani}}, \bibinfo {author} {\bibfnamefont {C.}~\bibnamefont {Schmid}}, \bibinfo {author} {\bibfnamefont {S.}~\bibnamefont {Spedicato}}, \bibinfo {author} {\bibfnamefont {S.}~\bibnamefont {Strohauer}}, \bibinfo {author} {\bibfnamefont {F.}~\bibnamefont {Wietschorke}}, \bibinfo {author} {\bibfnamefont {F.}~\bibnamefont {Flassig}}, \bibinfo {author} {\bibfnamefont {J.~J.}\ \bibnamefont {Finley}},\ and\ \bibinfo {author} {\bibfnamefont {K.}~\bibnamefont {M{\"u}ller}},\ }\bibfield  {title} {\bibinfo {title} {The dependence of timing jitter of superconducting nanowire single-photon detectors on the multi-layer sample design and slew rate},\ }\href {https://doi.org/10.1039/D2NR04494C} {\bibfield  {journal} {\bibinfo  {journal} {Nanoscale}\ }\textbf {\bibinfo {volume} {15}},\ \bibinfo {pages} {1086} (\bibinfo {year} {2023})}\BibitemShut {NoStop}%
\end{thebibliography}%


\begin{thebibliography}{4}%
\makeatletter
\providecommand \@ifxundefined [1]{%
 \@ifx{#1\undefined}
}%
\providecommand \@ifnum [1]{%
 \ifnum #1\expandafter \@firstoftwo
 \else \expandafter \@secondoftwo
 \fi
}%
\providecommand \@ifx [1]{%
 \ifx #1\expandafter \@firstoftwo
 \else \expandafter \@secondoftwo
 \fi
}%
\providecommand \natexlab [1]{#1}%
\providecommand \enquote  [1]{``#1''}%
\providecommand \bibnamefont  [1]{#1}%
\providecommand \bibfnamefont [1]{#1}%
\providecommand \citenamefont [1]{#1}%
\providecommand \href@noop [0]{\@secondoftwo}%
\providecommand \href [0]{\begingroup \@sanitize@url \@href}%
\providecommand \@href[1]{\@@startlink{#1}\@@href}%
\providecommand \@@href[1]{\endgroup#1\@@endlink}%
\providecommand \@sanitize@url [0]{\catcode `\\12\catcode `\$12\catcode `\&12\catcode `\#12\catcode `\^12\catcode `\_12\catcode `\%12\relax}%
\providecommand \@@startlink[1]{}%
\providecommand \@@endlink[0]{}%
\providecommand \url  [0]{\begingroup\@sanitize@url \@url }%
\providecommand \@url [1]{\endgroup\@href {#1}{\urlprefix }}%
\providecommand \urlprefix  [0]{URL }%
\providecommand \Eprint [0]{\href }%
\providecommand \doibase [0]{https://doi.org/}%
\providecommand \selectlanguage [0]{\@gobble}%
\providecommand \bibinfo  [0]{\@secondoftwo}%
\providecommand \bibfield  [0]{\@secondoftwo}%
\providecommand \translation [1]{[#1]}%
\providecommand \BibitemOpen [0]{}%
\providecommand \bibitemStop [0]{}%
\providecommand \bibitemNoStop [0]{.\EOS\space}%
\providecommand \EOS [0]{\spacefactor3000\relax}%
\providecommand \BibitemShut  [1]{\csname bibitem#1\endcsname}%
\let\auto@bib@innerbib\@empty
\bibitem [{\citenamefont {Ferrari}\ \emph {et~al.}(2018)\citenamefont {Ferrari}, \citenamefont {Schuck},\ and\ \citenamefont {Pernice}}]{Ferrari2018}%
  \BibitemOpen
  \bibfield  {author} {\bibinfo {author} {\bibfnamefont {S.}~\bibnamefont {Ferrari}}, \bibinfo {author} {\bibfnamefont {C.}~\bibnamefont {Schuck}},\ and\ \bibinfo {author} {\bibfnamefont {W.}~\bibnamefont {Pernice}},\ }\bibfield  {title} {\bibinfo {title} {Waveguide-integrated superconducting nanowire single-photon detectors},\ }\href {https://doi.org/10.1515/nanoph-2018-0059} {\bibfield  {journal} {\bibinfo  {journal} {Nanophotonics}\ }\textbf {\bibinfo {volume} {7}},\ \bibinfo {pages} {1725} (\bibinfo {year} {2018})}\BibitemShut {NoStop}%
\bibitem [{\citenamefont {Kerman}\ \emph {et~al.}(2006)\citenamefont {Kerman}, \citenamefont {Dauler}, \citenamefont {Keicher}, \citenamefont {Yang}, \citenamefont {Berggren}, \citenamefont {Gol'tsman},\ and\ \citenamefont {Voronov}}]{Kerman2006}%
  \BibitemOpen
  \bibfield  {author} {\bibinfo {author} {\bibfnamefont {A.~J.}\ \bibnamefont {Kerman}}, \bibinfo {author} {\bibfnamefont {E.~A.}\ \bibnamefont {Dauler}}, \bibinfo {author} {\bibfnamefont {W.~E.}\ \bibnamefont {Keicher}}, \bibinfo {author} {\bibfnamefont {J.~K.}\ \bibnamefont {Yang}}, \bibinfo {author} {\bibfnamefont {K.~K.}\ \bibnamefont {Berggren}}, \bibinfo {author} {\bibfnamefont {G.}~\bibnamefont {Gol'tsman}},\ and\ \bibinfo {author} {\bibfnamefont {B.}~\bibnamefont {Voronov}},\ }\bibfield  {title} {\bibinfo {title} {Kinetic-inductance-limited reset time of superconducting nanowire photon counters},\ }\href {https://doi.org/10.1063/1.2183810} {\bibfield  {journal} {\bibinfo  {journal} {Applied Physics Letters}\ }\textbf {\bibinfo {volume} {88}},\ \bibinfo {pages} {2} (\bibinfo {year} {2006})}\BibitemShut {NoStop}%
\bibitem [{\citenamefont {Strohauer}\ \emph {et~al.}(2023)\citenamefont {Strohauer}, \citenamefont {Wietschorke}, \citenamefont {Zugliani}, \citenamefont {Flaschmann}, \citenamefont {Schmid}, \citenamefont {Grotowski}, \citenamefont {M{\"u}ller}, \citenamefont {Jonas}, \citenamefont {Althammer}, \citenamefont {Gross}, \citenamefont {M{\"u}ller},\ and\ \citenamefont {Finley}}]{Strohauer2023}%
  \BibitemOpen
  \bibfield  {author} {\bibinfo {author} {\bibfnamefont {S.}~\bibnamefont {Strohauer}}, \bibinfo {author} {\bibfnamefont {F.}~\bibnamefont {Wietschorke}}, \bibinfo {author} {\bibfnamefont {L.}~\bibnamefont {Zugliani}}, \bibinfo {author} {\bibfnamefont {R.}~\bibnamefont {Flaschmann}}, \bibinfo {author} {\bibfnamefont {C.}~\bibnamefont {Schmid}}, \bibinfo {author} {\bibfnamefont {S.}~\bibnamefont {Grotowski}}, \bibinfo {author} {\bibfnamefont {M.}~\bibnamefont {M{\"u}ller}}, \bibinfo {author} {\bibfnamefont {B.}~\bibnamefont {Jonas}}, \bibinfo {author} {\bibfnamefont {M.}~\bibnamefont {Althammer}}, \bibinfo {author} {\bibfnamefont {R.}~\bibnamefont {Gross}}, \bibinfo {author} {\bibfnamefont {K.}~\bibnamefont {M{\"u}ller}},\ and\ \bibinfo {author} {\bibfnamefont {J.~J.}\ \bibnamefont {Finley}},\ }\bibfield  {title} {\bibinfo {title} {Site-{{Selective Enhancement}} of {{Superconducting Nanowire Single}}-{{Photon Detectors}} via {{Local Helium Ion Irradiation}}},\ }\href {https://doi.org/10.1002/qute.202300139}
  {\bibfield  {journal} {\bibinfo  {journal} {Advanced Quantum Technologies}\ }\textbf {\bibinfo {volume} {6}},\ \bibinfo {pages} {2300139} (\bibinfo {year} {2023})}\BibitemShut {NoStop}%
\bibitem [{\citenamefont {Kerman}\ \emph {et~al.}(2013)\citenamefont {Kerman}, \citenamefont {Rosenberg}, \citenamefont {Molnar},\ and\ \citenamefont {Dauler}}]{Kerman2013}%
  \BibitemOpen
  \bibfield  {author} {\bibinfo {author} {\bibfnamefont {A.~J.}\ \bibnamefont {Kerman}}, \bibinfo {author} {\bibfnamefont {D.}~\bibnamefont {Rosenberg}}, \bibinfo {author} {\bibfnamefont {R.~J.}\ \bibnamefont {Molnar}},\ and\ \bibinfo {author} {\bibfnamefont {E.~A.}\ \bibnamefont {Dauler}},\ }\bibfield  {title} {\bibinfo {title} {Readout of superconducting nanowire single-photon detectors at high count rates},\ }\href {https://doi.org/10.1063/1.4799397} {\bibfield  {journal} {\bibinfo  {journal} {Journal of Applied Physics}\ }\textbf {\bibinfo {volume} {113}},\ \bibinfo {pages} {144511} (\bibinfo {year} {2013})}\BibitemShut {NoStop}%
\end{thebibliography}%

\end{document}


\title{Supplementary:\\
Current-Crowding-Free Superconducting Nanowire
Single-Photon Detectors
}

\author{Stefan Strohauer}
\email{stefan.strohauer@tum.de}
\affiliation{Walter Schottky Institute, Technical University of Munich,
85748 Garching, Germany
}
\affiliation{TUM School of Natural Sciences, Technical University of Munich,
85748 Garching, Germany}
\author{Fabian Wietschorke}
\author{Christian Schmid}
\affiliation{Walter Schottky Institute, Technical University of Munich,
85748 Garching, Germany
}
\affiliation{TUM School of Computation, Information and Technology, Technical University of Munich,
80333 Munich, Germany
}
\author{Stefanie Grotowski}
\affiliation{Walter Schottky Institute, Technical University of Munich,
85748 Garching, Germany
}
\affiliation{TUM School of Natural Sciences, Technical University of Munich,
85748 Garching, Germany}
\author{Lucio Zugliani}
\affiliation{Walter Schottky Institute, Technical University of Munich,
85748 Garching, Germany
}
\affiliation{TUM School of Computation, Information and Technology, Technical University of Munich,
80333 Munich, Germany
}
\author{Björn Jonas}
\affiliation{Walter Schottky Institute, Technical University of Munich,
85748 Garching, Germany
}
\affiliation{TUM School of Computation, Information and Technology, Technical University of Munich,
80333 Munich, Germany
}
\author{Kai Müller}
\affiliation{Walter Schottky Institute, Technical University of Munich,
85748 Garching, Germany
}
\affiliation{TUM School of Computation, Information and Technology, Technical University of Munich,
80333 Munich, Germany
}
\affiliation{Munich Center for Quantum Science and Technology (MCQST),
80799 Munich, Germany}
\author{Jonathan J. Finley}
\email{jj.finley@tum.de}
\affiliation{Walter Schottky Institute, Technical University of Munich,
85748 Garching, Germany
}
\affiliation{TUM School of Natural Sciences, Technical University of Munich,
85748 Garching, Germany}
\affiliation{Munich Center for Quantum Science and Technology (MCQST),
80799 Munich, Germany}

\date{July 18, 2024}

\keywords{
current crowding,
superconducting thin film,
radiation damage,
He ion irradiation,
} %

\maketitle

\section{Detection pulse characterization}
In this section, we analyze the recovery time and pulse height of a detection pulse after photon absorption, since both are important performance parameters of single-photon detectors. 

\begin{figure}
 \centering
 \includegraphics{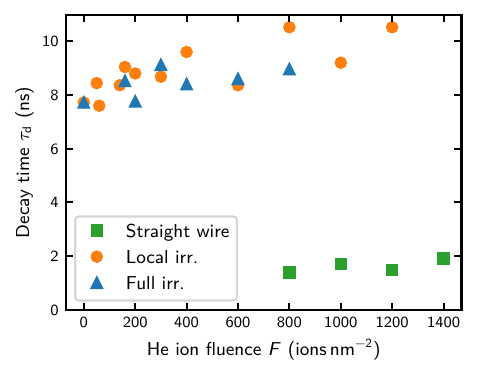}
 \caption{Decay time \vs He ion fluence for locally irradiated \sspds together with fully irradiated \sspds and straight wires as reference devices.}
\label{fig:fall_time_vs_dose}
\end{figure}

The recovery time determines the detector's maximum count rate and can be estimated from the time constant $\tau_\mathrm{d}$ of the exponential decay of a detection voltage pulse \cite{Ferrari2018, Kerman2006}.
As shown in \cref{fig:fall_time_vs_dose}, the decay time of all device types increases with increasing He ion fluence in agreement with previous measurements \cite{Strohauer2023}.
At the same time, the decay time of locally irradiated \sspds follows the same curve as that of fully irradiated \sspds, while pulses of straight wires show a significantly faster decay due their smaller kinetic inductance.
Thus, local irradiation shows no drawback with respect to the decay time compared to full irradiation of \sspds.

\begin{figure}
 \centering
 \includegraphics{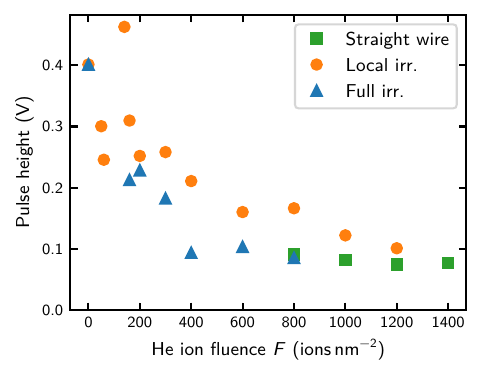}
 \caption{Pulse height \vs He ion fluence for locally irradiated \sspds together with fully irradiated \sspds and straight wires as reference devices.}
\label{fig:pulse_height_vs_dose}
\end{figure}

As shown in \cref{fig:pulse_height_vs_dose} the pulse height of locally irradiated \sspds is higher than that of fully irradiated \sspds due to higher available bias currents after irradiating only locally.
This is beneficial since higher detection voltage pulses are easier to process for the readout electronics.


\section{Saturation plateau width of locally irradiated \sspds}
\begin{figure}
 \centering
 \includegraphics{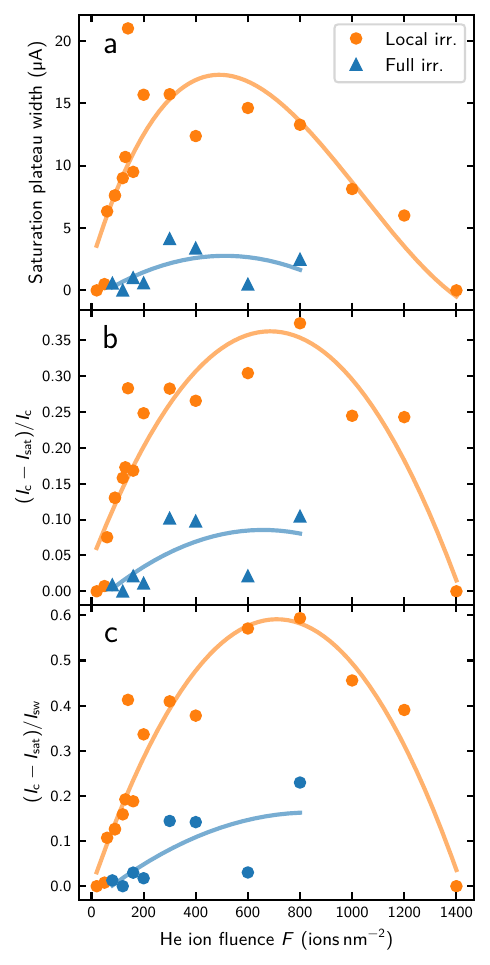}
 \caption{Count rate saturation plateau width of locally and fully irradiated \sspds \vs He ion fluence. \textbf{a}, Absolute saturation plateau width. \textbf{b}, Relative saturation plateau width as given in \cref{eq:RSPW}. \textbf{c}, Saturation plateau width, normalized to the switching current of the corresponding devices measured without shunt resistor.}
\label{fig:Composition_SPW_RSPW_RSPWIsw}
\end{figure}

The relative saturation plateau width is given by
\begin{equation}
    \sigma_\mathrm{rel} = 
    \frac{I_\mathrm{c} - I_\mathrm{sat}}{I_\mathrm{c}}
    \;,
    \label{eq:RSPW}
\end{equation}
with the absolute saturation plateau size given by the difference between the critical current $I_\mathrm{c}$ and the current $I_\mathrm{sat}$ where the saturation plateau begins.
Since we use a shunt resistor for the CR measurements to prevent latching of the detectors (\qty{20.4}{\ohm} at \qty{1}{\K}), the \sspds transition to the relaxation oscillation regime at $I_\mathrm{c}$ before switching to the latching state.
In this regime the \sspd emits a periodic train of voltage pulses and the average voltage drop across the \sspd increases with increasing bias current \cite{Kerman2013}.
\cref{fig:Composition_SPW_RSPW_RSPWIsw}a shows the absolute saturation plateau width $I_\mathrm{c} - I_\mathrm{sat}$ where a steep increase in saturation plateau width is observed for He ion fluences up to \qty{200}{\ipsn}, followed by a decrease beyond \qty{600}{\ipsn}. 
At the same time, the critical current decreases with increasing He ion fluence.
As shown in \cref{fig:Composition_SPW_RSPW_RSPWIsw}b, the resulting relative saturation plateau width increases for small He ion fluences, peaks between \qty{600}{\ipsn} and \qty{1000}{\ipsn} before it decreases again for higher fluences.
This relation is even better visible in \cref{fig:Composition_SPW_RSPW_RSPWIsw}c after normalizing the saturation plateau width to the switching current of the devices without shunt resistor instead of the critical current during CR measurements.

\bibliography{HIM_paper_II_supplementary}%